\documentclass[draft]{agujournal2019}
\usepackage{url, amsmath, amsfonts, booktabs} 
\usepackage{subfiles}
\usepackage{lineno}
\usepackage[inline]{trackchanges} 
\usepackage{soul}

\newcommand*\patchAmsMathEnvironmentForLineno[1]{
  \expandafter\let\csname old#1\expandafter\endcsname\csname #1\endcsname
  \expandafter\let\csname oldend#1\expandafter\endcsname\csname end#1\endcsname
  \renewenvironment{#1}
  {\linenomath\csname old#1\endcsname}
  {\csname oldend#1\endcsname\endlinenomath}}
  \newcommand*\patchBothAmsMathEnvironmentsForLineno[1]{
  \patchAmsMathEnvironmentForLineno{#1}
  \patchAmsMathEnvironmentForLineno{#1*}}
  \AtBeginDocument{
  \patchBothAmsMathEnvironmentsForLineno{equation}
  \patchBothAmsMathEnvironmentsForLineno{align}
  \patchBothAmsMathEnvironmentsForLineno{flalign}
  \patchBothAmsMathEnvironmentsForLineno{alignat}
  \patchBothAmsMathEnvironmentsForLineno{gather}
  \patchBothAmsMathEnvironmentsForLineno{multline}
}

\draftfalse

\journalname{Geophysical Research Letters}

\begin{document}

%
%

\title{Calibration of a neural network ocean closure for improved mean state and variability}

%
%



\authors{Pavel Perezhogin\affil{1}, Alistair Adcroft\affil{2}, Laure Zanna\affil{1}}

\affiliation{1}{Courant Institute School of Mathematics, Computing, and Data Science, New York University, New York, NY, USA}

\affiliation{2}{Program in Atmospheric and Oceanic Sciences, Princeton University, Princeton, NJ, USA}




\correspondingauthor{Pavel Perezhogin}{pp2681@nyu.edu}





\begin{keypoints}
\item Calibrated data-driven eddy parameterization reduces the mean state and variability errors by factors of $1.7-3.3$ in coarse ocean models
\item Ensemble Kalman Inversion effectively optimizes neural network parameters in the presence of chaotic ocean dynamics
\item Efficient calibration is achievable with short simulations without integrating the ocean model to statistical equilibrium
\end{keypoints}


%
%

%
%

%
\begin{abstract}
Global ocean models exhibit biases in the mean state and variability, particularly at coarse resolution, where mesoscale eddies are unresolved. To address these biases, parameterization coefficients are typically tuned ad hoc. Here, we formulate parameter tuning as a calibration problem using Ensemble Kalman Inversion (EKI). We optimize parameters of a neural network parameterization of mesoscale eddies in two idealized ocean models at coarse resolution. The calibrated parameterization reduces errors by factors of $1.7-3.3$ in the time-averaged fluid interfaces and their variability compared to the unparameterized model, depending on the metric and configuration. The EKI method is robust to noise in time-averaged statistics arising from chaotic ocean dynamics. Furthermore, we propose an efficient calibration protocol that bypasses integration to statistical equilibrium by carefully choosing an initial condition. 
These results demonstrate that systematic calibration can substantially improve coarse-resolution ocean simulations and provide a practical pathway for reducing biases in global ocean models.
\end{abstract}

\section*{Plain Language Summary}
Ocean models used for climate prediction have persistent errors because they cannot capture small-scale swirling currents called eddies. Models often include mathematical corrections called parameterizations to approximate the effects of these missing eddies, but the adjustable settings in these corrections are usually chosen by hand through trial and error. We use a machine learning approach combined with an automatic tuning method to identify improved settings for an eddy parameterization in two simplified ocean simulations. Our tuning method reduces errors in both the average ocean state and its natural fluctuations by roughly half compared to an untuned model. Importantly, the method works well even when ocean statistics are noisy due to the chaotic nature of ocean currents, and it can be applied with relatively short simulations rather than waiting hundreds of years of simulations for the ocean model to fully adjust. These results offer a practical path toward reducing longstanding biases in the ocean models used for climate projections.

%
%

%


%
%
%
%

\section{Introduction}

Global ocean models exhibit substantial biases in the mean state and variability \cite{wang2014global, richter2020overview}. For example, representing variability of the Western Boundary Currents (WBC) is challenging across a wide range of horizontal resolutions of the ocean models, including non-eddy-resolving \cite{grooms2024stochastic}, eddy-permitting \cite{juricke2020ocean}, and submesoscale-permitting \cite{uchida2022cloud}. The mean-state biases include errors at the air-sea interface and subsurface isopycnal structure \cite{griffies2015impacts, adcroft2019gfdl}. These biases are often mitigated by incorporating various parameterizations \cite{andrejczuk2016oceanic, juricke2017stochastic, juricke2020ocean, chang2023remote, grooms2024stochastic}. However, the adjustment of parameterization coefficients is frequently performed in an ad hoc manner. Here, we formalize this process as a calibration problem, see also \citeA{cooper2015optimisation, cooper2017optimisation}.

Recently, multiple data-driven eddy parameterizations have been proposed to reduce biases in the ocean mean state and variability \cite{zanna2020data, guillaumin2021stochastic, zhang2023implementation, perezhogin2025generalizable, kamm2026assessing}. These parameterizations perform well at an eddy-permitting resolution ($1/4^{\circ}$). However, their performance often degrades at a coarser resolution ($1/2^{\circ}$), a particularly challenging resolution for testing eddy parameterizations \cite{jansen2019toward, yankovsky2024}. 
Developing skillful data-driven parameterizations for coarse ocean models is especially important, as these models are widely used in climate simulations \cite{grooms2024stochastic} and offer the greatest potential for bias reduction \cite{perezhogin2023generative}. We build on recent works demonstrating that optimizing the parameterization parameters in online simulations can significantly improve the model fidelity at coarse resolution \cite{kochkov2021machine, frezat2022posteriori, lopez2022training, ouala2024online, maddison2024online, christopoulos2024online, wagner2025formulation, yan2025adjoint, lee2025nori}. In contrast to physics-based parameterizations, machine-learned parameterizations often contain too many parameters for manual tuning, motivating the use of automatic calibration methods. 

Calibration methods are designed to automatically adjust the free parameters of a parameterization by minimizing the mismatch between the output of the coarse-resolution ocean model and observations, i.e., by minimizing a prescribed loss function. Calibration of ocean models is exceptionally expensive due to the long spin-up period, which can take hundreds of years \cite{williamson2017tuning, mrozowska2025bayesian}, as well as the extended time window required to accurately estimate time-averaged statistics. In this work, we demonstrate the robustness of the Ensemble Kalman Inversion \cite<EKI, >{iglesias2013ensemble} calibration method to noise in temporal averages arising from the ocean's chaotic dynamics. Furthermore, we propose a simple method for calibrating fast physical processes without integrating the ocean model to statistical equilibrium, a longstanding challenge in climate modeling \cite{delsole2024tuning}.

Our goal is to determine to what extent calibration of a neural-network eddy parameterization by \citeA{perezhogin2025generalizable} can improve the mean state and variability of an idealized GFDL MOM6 ocean model \cite{adcroft2019gfdl} at coarse resolution ($1/2^{\circ}$). We constrain the neural network with physical equivariances to enhance generalization \cite{kashinath2021physics} and reduce the number of calibrated parameters. We consider two idealized ocean configurations, which serve different purposes. The simplest configuration is used to assess the convergence of the calibration algorithm and its robustness to the noise \cite{dunbar2021calibration, dunbar2022ensemble, howland2022parameter, gjini2025ensemble}. A more complex configuration is used to demonstrate the applicability of our calibration protocol to ocean models with long spin-up times.


\section{Methods}

\begin{figure}[h!]
\centering
\includegraphics[scale=0.3]{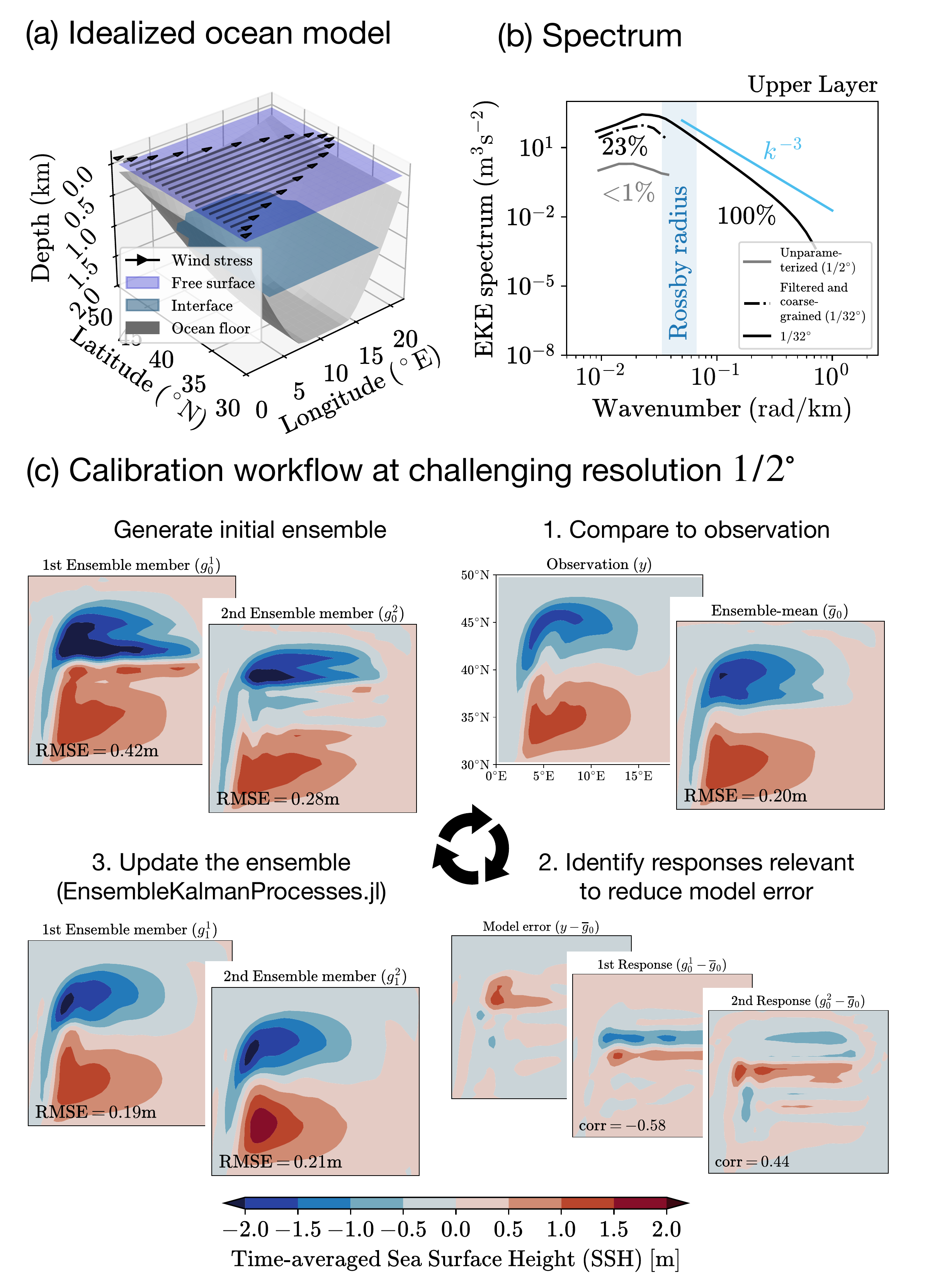}
\caption{(a) Idealized wind-driven ocean model GFDL MOM6 in a double-gyre configuration. (b) The eddy kinetic energy (EKE) spectrum as a function of isotropic horizontal wavenumber in the upper fluid layer and domain $5^{\circ}\mathrm{E}$$-$$15^{\circ}\mathrm{E}$ $\times $ $35^{\circ}\mathrm{N}$$-$$45^{\circ}\mathrm{N}$. The percentages represent the integral over the spectrum relative to that of the high-resolution simulation. Panel (c) shows how the Ensemble Kalman Inversion calibration algorithm interacts with the coarse ocean model in order to update the free parameters of the parameterization such that the time-mean sea surface height is as close as possible to the filtered and coarse-grained high-resolution simulation ("observation").}
\label{fig:calibation_workflow}
\end{figure}

In this section, we describe how improving the mean state and variability of the ocean can be framed as a calibration problem. We introduce idealized ocean models and eddy parameterization, followed by the choice of the calibration method, loss function, and a method to avoid long spin-up. A simplified workflow is illustrated in Figure \ref{fig:calibation_workflow}.


\subsection{Idealized ocean models} \label{sec:ocean_model}

We consider two idealized configurations of the GFDL MOM6 ocean model: Double Gyre \cite<DG, >{zhang2023implementation, perezhogin2024stable, zhang2025weno} and NeverWorld2 \cite<NW2, >{marques2022neverworld2, yankovsky2022influences, yankovsky2024}. Both configurations represent adiabatic ocean dynamics and solve the stacked shallow water equations, with the circulation driven by prescribed wind stress. Our goal is to improve the time-averaged statistical properties of the ocean circulation in the parameterized simulation at a coarse horizontal resolution $1/2^{\circ}$, and bring them closer to the statistics of the filtered and coarse-grained high-resolution simulation at resolution $1/32^{\circ}$. All simulations use the biharmonic Smagorinsky eddy viscosity with coefficient $C_S=0.06$ \cite{adcroft2019gfdl} to maintain numerical stability. 

The DG configuration (Figure \ref{fig:calibation_workflow}(a)) represents a midlatitude basin with two fluid layers. The imposed wind stress drives two counter-rotating gyres separated by an eastward jet, serving as an idealized model of western boundary current systems. The target ocean circulation is obtained by integrating the $1/32^{\circ}$ model for 100 years from rest, discarding the first 10 years for spin-up. The coarse ($1/2^{\circ}$) parameterized model is also evaluated in 100-year simulations. However, during calibration, the coarse model is integrated only for 20 years. 

The NW2 configuration represents an idealized Atlantic sector model with 15 fluid layers, featuring multiple circulation regimes, including a circumpolar current in an idealized Southern Ocean, midlatitude gyres, and equatorial flows. The high-resolution simulation ($1/32^{\circ}$) was spun up in multiple stages in \citeA{marques2022neverworld2}. The coarse ocean model at $1/2^{\circ}$ resolution is integrated for 5 years during the calibration stage and for 30000 days from rest during evaluation of the calibrated parameterization. In all simulations, we analyze the final 800 days. 


\subsection{Neural-network parameterization of mesoscale eddies} \label{sec:parameterization}
We use the recently developed data-driven parameterization of mesoscale eddies by \citeA{perezhogin2025generalizable}, which modifies the  horizontal momentum balance equation:
\begin{equation}
     \partial_t \mathbf{u} =\cdots + \nabla \cdot \mathbf{T},
\end{equation}
where $\mathbf{T} \in \mathbb{R}^{2\times2}$ is the horizontal stress tensor predicted by the parameterization, $\mathbf{u} = (u,v)$ is the vector of filtered horizontal velocities, and $\nabla=(\partial_x, \partial_y)$. This parameterization represents the inverse kinetic energy cascade across the grid scale ($\mathbf{u} \cdot (\nabla \cdot \mathbf{T}) > 0$ on average, \citeA{storer2023global}) known as backscatter \cite{kraichnan1976eddy, chasnov1991simulation, frederiksen1997eddy, berner2009spectral, jansen2014parameterizing, juricke2019ocean}.

The parameterized stress tensor is predicted as follows
\begin{equation}
        \mathbf{T}(\mathbf{X}, \Delta) = \gamma \Delta^2 ||\mathbf{X}||_2^2 f_{\phi} (\mathbf{X} / ||\mathbf{X}||_2), \label{eq:ann_eq}
\end{equation}
where $\Delta$ is the coarse grid spacing, $\gamma$ is a scaling coefficient, which is equal to $1$ during training, $f_{\phi}$ is the Artificial Neural Network (ANN) with free parameters $\phi$. The vector of input features, $\mathbf{X} \in \mathbb{R}^{27}$, consists of horizontal velocity gradients (${\sigma}_S = \partial_y {u} + \partial_x {v}$, ${\sigma}_D = \partial_x {u} - \partial_y {v}$, ${\omega}  = \partial_x {v} - \partial_y {u}$) evaluated on a $3 \times 3$ spatial stencil. Normalization of input features with $||\mathbf{X}||_2$ and output features with $\Delta^2 ||\mathbf{X}||_2^2$ introduces dimensional consistency and was shown to improve generalization to out-of-distribution data \cite{perezhogin2025generalizable}.  

The parameterization (Eq. \eqref{eq:ann_eq}) encapsulates many hard constraints that preserve physical invariances, including Galilean invariance, dimensional consistency, and conservation of momentum and angular momentum. There were only two implemented soft constraints in \citeA{perezhogin2025generalizable} -- rotational and reflectional invariances. This approach is reasonable when the parameterization is trained offline. However, online recalibration can violate invariances during optimization of the online loss function. Thus, in this work, we further constrain the parameterization (Eq. \eqref{eq:ann_eq}) and implement rotational and reflectional invariances as hard constraints following \citeA{guan2022learning, pawar2023frame, connolly2025deep} (see Text S1-S3). 
We will refer to the new parameterization as eANN (equivariant ANN).

We train the eANN on the global ocean dataset at coarse resolutions in a range of $0.4^{\circ}-1.5^{\circ}$ using the same algorithm as in \citeA{perezhogin2025generalizable} and report similar offline performance (see Figure S1 in SI). We note that the set of resolutions covered in the training dataset is well-suited to the online simulations we consider here ($1/2^{\circ}$). 

\subsection{Ensemble Kalman Inversion} \label{sec:calibration}
The goal of calibration is to adjust the tunable parameters of the parameterization so that the statistics of the coarse ocean model are as close as possible to those of the filtered, coarse-grained high-resolution data. This problem can be framed as the minimization of the following loss function \cite{gjini2025ensemble}:
\begin{equation}
    \mathcal{L}(\theta) = || R^{-1/2} (y - \mathcal{G}(\theta)) ||_2^2. \label{eq:opt_problem}
\end{equation}
Here, $\theta \in \mathbb{R}^{n_p}$ is a vector of length $n_p$ representing parameters to be calibrated, $y  \in \mathbb{R}^{n_o}$ is a vector of observations of length $n_o$ (i.e., in our case, statistics of filtered and coarsened high-resolution simulation), $\mathcal{G}(\theta)$ is a forward model evaluation (i.e., statistics of the coarse ocean simulation, performed at a given set of parameters $\theta$), and $R$ is the covariance matrix of the observational error.

We solve the optimization problem (Eq. \eqref{eq:opt_problem}) using a gradient-free optimization method -- Ensemble Kalman Inversion \cite{iglesias2013ensemble} implemented in the software package \texttt{Ensemble\-Kalman\-Processes.jl} \cite{oliver2022ensemblekalmanprocesses}.
A particular method of Kalman inversion we use among the methods implemented in the package is the Ensemble Transform Kalman Inversion \cite<ETKI,>{huang2022efficient}. This choice is made for two reasons: the possibility to explore the parameter space (the ensemble size is not tightly fixed to the number of parameters) and scalability with respect to the observational dimension ($n_o$). 

The ETKI method is initialized by sampling an ensemble of $n_e$ parameter vectors, denoted by $\theta_0^1$, ..., $\theta_0^{n_e}$. 
The forward model  is evaluated at these parameter vectors, giving $g_0^1=G(
\theta_0^1)$, ..., $g_0^{n_e}=G(
\theta_0^{n_e})$. Here, the superscript indexes the ensemble members, while the subscript denotes the iteration number, where 0 corresponds to the initial ensemble. The ensemble is then updated over multiple iterations, as described below (see Figure \ref{fig:calibation_workflow}(c) for illustration).

At each iteration, the ensemble-mean parameter vector is updated as follows \cite{gjini2025ensemble}:
\begin{equation}
    \overline{\theta}_{j+1} = \overline{\theta}_{j} + \delta t 
    \Theta_j (I + \delta t G_j^T R^{-1} G_j)^{-1} G_j^T R^{-1} (y - \overline{g}_j), \label{eq:ETKI_ensemble_mean}
\end{equation}
where $\delta t > 0$ is the scheduler step \cite{iglesias2021adaptive}, $j$ is the iteration number, $I$ is the identity matrix, $\overline{\theta}_j = 1/n_e\sum_i \theta_j^i$ is the ensemble-mean parameter vector, $\overline{g}_j = 1/n_e\sum_i g_j^i$ is the ensemble-mean forward model evaluation, $\Theta_j$ and $G_j$ are normalized perturbation matrices:
\begin{gather}
    \Theta_j=\frac{1}{\sqrt{n_e-1}} \left(\theta_j^1 -\overline{\theta}_j, ..., \theta_j^{n_e} -\overline{\theta}_j \right) \in \mathbb{R}^{n_p \times n_e}, \\
    G_j=\frac{1}{\sqrt{n_e-1}} \left(g_j^1 -\overline{g}_j, ..., g_j^{n_e} -\overline{g}_j \right) \in \mathbb{R}^{n_o \times n_e}.
\end{gather}

The mechanism of the update equation \eqref{eq:ETKI_ensemble_mean}  is as follows. We first compare the ensemble-mean prediction $\overline{g}_j$ with the observational vector $y$ (panel 1 in Figure \ref{fig:calibation_workflow}(c)). Next, we identify the individual responses that project onto the model error, i.e. $G_j^T R^{-1} (y - \overline{g}_j) \neq 0$ (see panel 2 in Figure \ref{fig:calibation_workflow}(c)). Finally, we update the ensemble-mean parameter vector $\overline{\theta}_j$ along a direction that, on average, reduces the model error. The described mechanism is similar to gradient descent using the ensemble-based approximation of the gradient and preconditioning \cite{chada2020iterative, vernon2025nesterov}.

The update of the ensemble mean (Eq. \eqref{eq:ETKI_ensemble_mean}) is followed by the update of the perturbations:
\begin{equation}
    \Theta_{j+1} = \Theta_j (I + \delta t G_j^T R^{-1} G_j)^{-1/2},
\end{equation}
which shrinks the ensemble to the consensus.

\subsection{Loss function and parameters} \label{sec:loss}
Our goal is to improve both the mean and variability of the interfaces between fluid layers, as these are often used to assess the impact of eddy parameterizations \cite{juricke2020ocean, grooms2024stochastic, yankovsky2024, balwada2025design}.  Fluid interfaces, unlike other spatial fields, are associated with large-scale horizontal circulation patterns (i.e., the streamfunction; \citeA{vallis2017atmospheric}), which are well resolved on the coarse grid.

The observational vector $y$ consists of the time-averaged interfaces (denoted by $\overline{\eta}^t$) and their temporal standard deviation (denoted by $\sqrt{\overline{\eta'^2}^t}$):
\begin{equation}
    y = \begin{bmatrix}
        \overline{{\eta}}^t \\
        \sqrt{\overline{{\eta}'^2}^t}
    \end{bmatrix}, \label{eq:obs_vector}
\end{equation}
where ${\eta}$ is the flattened 3D array of the normalized vertical coordinate of the fluid interfaces (for normalization, see Text S4 in SI). The output from filtered and coarse-grained high-resolution simulation (vector $y$) and the output of the coarse ocean model (vectors $g_{j}^i$) are processed in the same manner. 

We specify the simplest observational noise model ($R = I$), where $I$ is the identity matrix, so as not to alter the inner product and the loss function, which already contains normalized variables:
\begin{equation}
        \mathcal{L}(\theta) = ||y - \overline{g}_j||_2^2. \label{eq:loss_function}
\end{equation}

The loss function (Eq. \eqref{eq:loss_function}) represents a multi-objective optimization problem, in which we intentionally chose equal weights for the time-averaged and the standard deviations of the interfaces. That way, we assign equal contributions for the potential energy bias in representation of the mean state and eddies, which are proportional to $(\overline{\eta}^t)^2$ and $\overline{\eta'^2}^t$, respectively. Energy-based $l_2$ norm is a popular choice in the analysis of model errors \cite{tuppi2023simultaneous} and optimal disturbances \cite{zasko2023optimal}. 
Unlike \citeA{yankovsky2024} and \citeA{pudig2025parameterizing}, we exclude domain-integrated metrics from the calibration protocol as they provide little information about the spatial patterns of the mean state and variability. Additionally, domain-integrated metrics may disrupt the optimization loss dynamics because the calibration algorithm struggles to strike a balance between errors in spatial fields and scalar fields.

We consider the weights and biases of the eANN in the last layer, as well as a coefficient in front of the parameterization $\gamma$, as a vector of tunable parameters $\theta$ of size $n_p=\mathrm{dim}(\theta)=14$. Restricting calibration to the deepest layers of a neural network is a common practice \cite{pahlavan2024explainable}. Furthermore, the first layer is responsible for feature extraction \cite{guan2022stable}, and calibrating it can significantly degrade the eANN’s generalization ability.  We note that the number of parameters was considerably reduced by using rotational and reflection equivariances as hard constraints (see Text S2 and Table S1 in SI). For constructing an initial ensemble, we perturb the offline trained values of parameter vector $\theta$ by $25 \%$ of their magnitude (see Table S2 in SI). Note that manually tuning 14 parameters via grid search is challenging, as it would require approximately $10^{14}$ simulations. 

Once the calibration problem is defined, the ETKI update rule (Eq. \eqref{eq:ETKI_ensemble_mean}) depends on two parameters: the ensemble size ($n_e$) and scheduler step ($\delta t$). We use $n_e=100$ in the DG configuration, following the recommendation of \citeA{oliver_dunbar_2025_16616326} for our parameter dimension. In the NW2 configuration, the ensemble size is doubled because coarse ocean simulations blow up for a bigger percentage of the ensemble members. The scheduler step controls the tradeoff: larger values accelerate the convergence, but the ensemble spread collapses faster. We choose fixed values of $\delta t$ manually ($\delta t=2$ in DG and $\delta t=0.025$ in NW2) to ensure monotonic decrease of the loss function (Eq. \eqref{eq:loss_function}) while avoiding excessively rapid ensemble collapse. The sensitivity to $\delta t$ and the random seed is shown in Figure S2 in SI.

\subsection{Calibrating an ocean model without integrating to statistical equilibrium} \label{sec:method}

Calibration in the NW2 configuration is computationally demanding, as a single simulated day requires 400 times as many CPU-core hours as in the DG configuration. Moreover, the spin-up time from rest is approximately 100 years in NW2, compared with approximately 5 years in DG. Consequently, we developed a method that reduces the computational cost of calibration in the NW2  configuration by using short simulations (5 years):
\begin{itemize}
    \item We initialize the coarse ocean model with the filtered and coarse-grained snapshot from the spun-up high-resolution simulation,
    \item We allow the coarse ocean model to adjust and partially forget the initial eddy field during a short spin-up (1000 days), and compute the loss function by averaging the statistics over an additional 800 days.
\end{itemize}
Here, the time-averaging interval (800 days) controls the amount of noise in the evaluation of the loss function (Eq. \eqref{eq:loss_function}). We found that considerably reducing the time-averaging interval degrades the performance of the calibrated parameterization (Figure S5 in SI). On the contrary, reducing the spin-up time (1000 days) to zero does not degrade performance (Figure S5 in the SI). This likely happened because, contrary to other studies on ocean model calibration, we use the exact initial condition, which allows us to evaluate the loss function without waiting for the ocean model to adjust. We suggest that the simulation length and its split into spin-up and time-averaging intervals are hyperparameters that should be optimized in future applications.


\section{Results}
We consider recalibration of the eddy parameterization in the coarse ocean model at a challenging resolution of $1/2^{\circ}$, which resolves less than 25\% of the eddy kinetic energy in the DG configuration in the filtered and coarse-grained high-resolution model, see Figure \ref{fig:calibation_workflow}(b).
First, we evaluate the efficiency of the calibration algorithm in a simple DG configuration and demonstrate its robustness to noise. Second, we apply the calibration algorithm to a substantially more expensive NW2 configuration and show that long-time statistics can be improved using only short simulations for calibration.

\subsection{Simple ocean configuration}

\begin{figure*}[h!]
\centering
\includegraphics[scale=0.5]{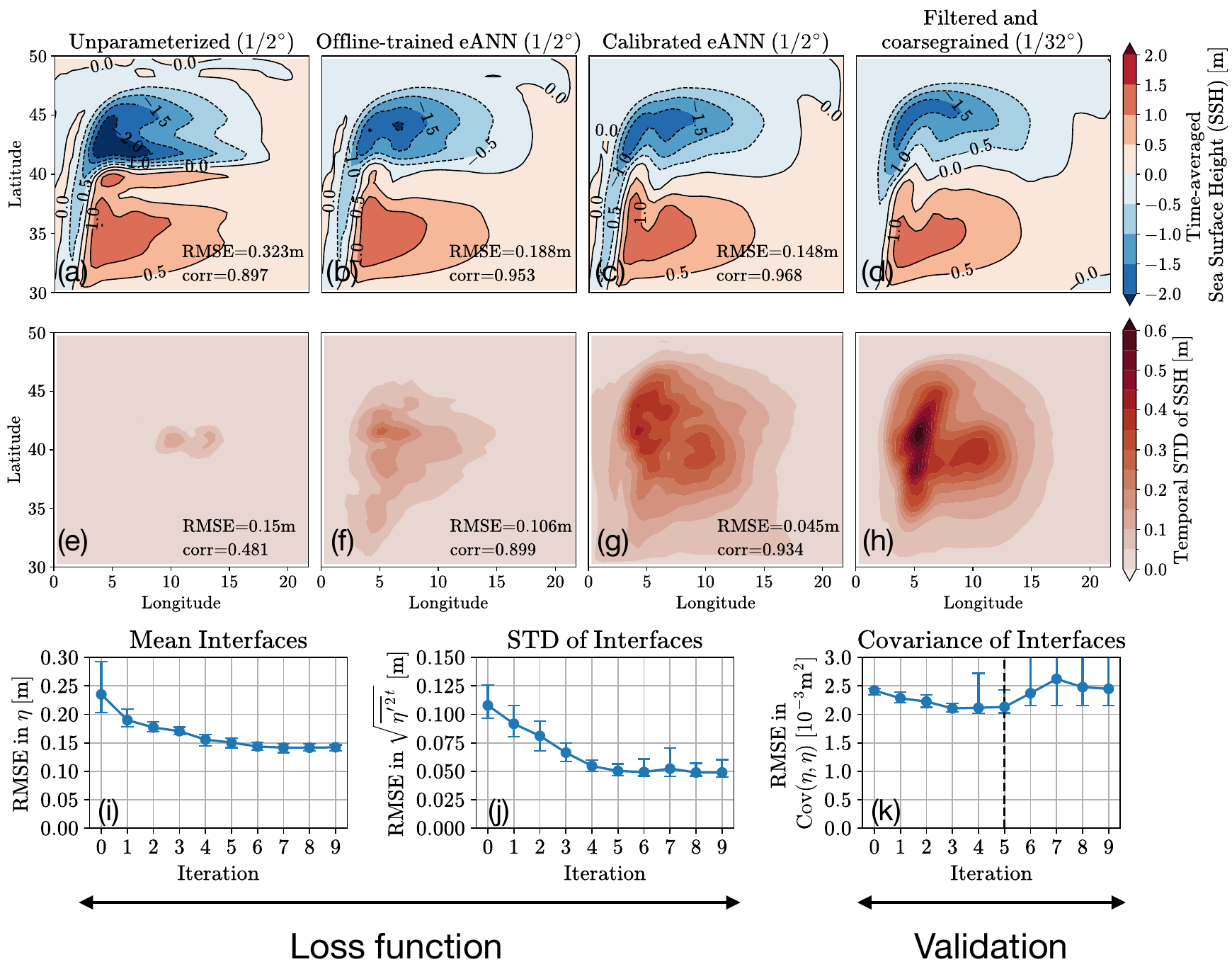}
\caption{Calibration of the eddy parameterization (eANN) in Double Gyre configuration. The upper row shows time-averaged sea surface height (SSH), and the second row shows the temporal standard deviation of SSH. On these panels, all simulations are 100 years long and results are averaged over 90 years. We show coarse ocean models ($1/2^{\circ}$) with: (a,e) no parameterization,
(b,f) parameterization trained offline with manually adjusted scaling coefficient ($\gamma=1.4$),
(c,g) calibrated parameterization. Panels (d,h) show the filtered and coarse-grained high-resolution simulation ($1/32^{\circ}$). The lowest row shows the convergence of the calibration process, as assessed by two metrics included in the loss function (i, j) and one metric excluded from the loss function (k). Blue markers show the ensemble median, and error bars show the 25\% and 75\% quantiles.
}
\label{fig:convergence}
\end{figure*}


Our goal is to show that calibrating a sufficiently expressive parameterization, such as a neural network, can significantly improve the mean state and variability of the ocean model. 
For this purpose, we run the calibration algorithm described in Section \ref{sec:calibration} for 10 iterations. At each iteration, this algorithm runs an ensemble of coarse online simulations in the DG configuration described in Section \ref{sec:ocean_model},
updates the weights and biases of the parameterization described in Section \ref{sec:parameterization}, attempting to minimize the loss function described in Section \ref{sec:loss}.

The coarse unparameterized model has a strong bias in the time-averaged sea surface height (SSH) with Root Mean Squared Error (RMSE) equal to 0.323m, and too low variability (RMSE in SSH standard deviation (STD) is 0.15m), see Figure \ref{fig:convergence} (a,e). 

We first isolate the effect of tuning only the scaling coefficient $\gamma$ in front of the offline-trained eANN parameterization, while keeping the neural network's weights and biases unchanged. The optimal coefficient $\gamma=1.4$ is determined as described in Figure S3 in SI. The offline-trained eANN with tuned coefficients reduces biases in the mean state and variability by $41.8\%$ and $29.3\%$, respectively, compared to the unparameterized model; see Figure \ref{fig:convergence} (b,f).

Second, we show the effect of jointly calibrating the scaling coefficient and the last layer of the neural network. The calibrated eANN parameterization reduces biases in the mean state and variability by $21.3\%$ and $57.5\%$, respectively, compared to the offline-trained parameterization with optimal scaling coefficient (Figure \ref{fig:convergence} (c,g)). The spatial pattern of the SSH standard deviation is accurately reproduced in the parameterized model, both near the boundary (Longitude = 5) and in the boundary current extension (Longitude = 10). This is especially important for future applications of our approach, as reproducing the variability in WBC extension is a challenging problem in global ocean models \cite{juricke2020ocean, uchida2022cloud}.

The statistics of the simulations used for calibration are affected by noise arising from a relatively short time-averaging interval (10 years) compared to the evaluation runs (90 years); see Figure S4 in the SI. Nevertheless, the calibration method is robust to noise, as evidenced by the monotonic decrease of the loss function (Figure \ref{fig:convergence} (i,j)).

The calibration algorithm schedules an ensemble of 100 ocean simulations at every iteration. Thus, over 10 iterations, we have 1000 simulations with different parameter vectors. Here, we describe how we chose one simulation shown in Figure \ref{fig:convergence} (c,g). Our parameterization is strongly constrained by physics and thus cannot compensate for all numerical model errors. As a result, we face not only parametric uncertainty but also model-form uncertainty (structural error) \cite{williamson2017tuning, prevost2025detection, shin2026accelerated}.
In such a setting, an optimal set of parameter values depends considerably on the choice of the loss function, which determines which biases to prioritize compensating. In this work, we chose a simple loss function (see Section \ref{sec:loss}) after extensive experimentation. Due to structural errors, minimizing the loss function may cause other physical metrics to deteriorate. Here, we show that the RMSE in the covariance matrix of interfaces is improving over the first 3 iterations, then remains on a plateau until the 5th iteration, and after that starts growing (Figure \ref{fig:convergence} (k)). Consequently, a further small improvement of the loss function is possible, but at the expense of generating strong unphysical modes of variability. This phenomenon in model tuning is known as overfitting \cite{williamson2017tuning}. We use the described validation metric to implement early stopping. That is, we consider the parameterization from the 5th iteration to be the best-performing one. Furthermore, we average the parameter vector over an ensemble to select a parameterization shown in panels (c,g) in Figure \ref{fig:convergence}. 
We note that the need for validation here may stem from the optimization being underconstrained: we consider only one simple ocean configuration. Thus, seeking a single parameter set suitable for multiple flow regimes \cite{lopez2022training, wagner2025formulation} might eliminate the need for a validation procedure, as we show in the next section.

\begin{figure*}[t!]
\centering
\includegraphics[scale=0.5]{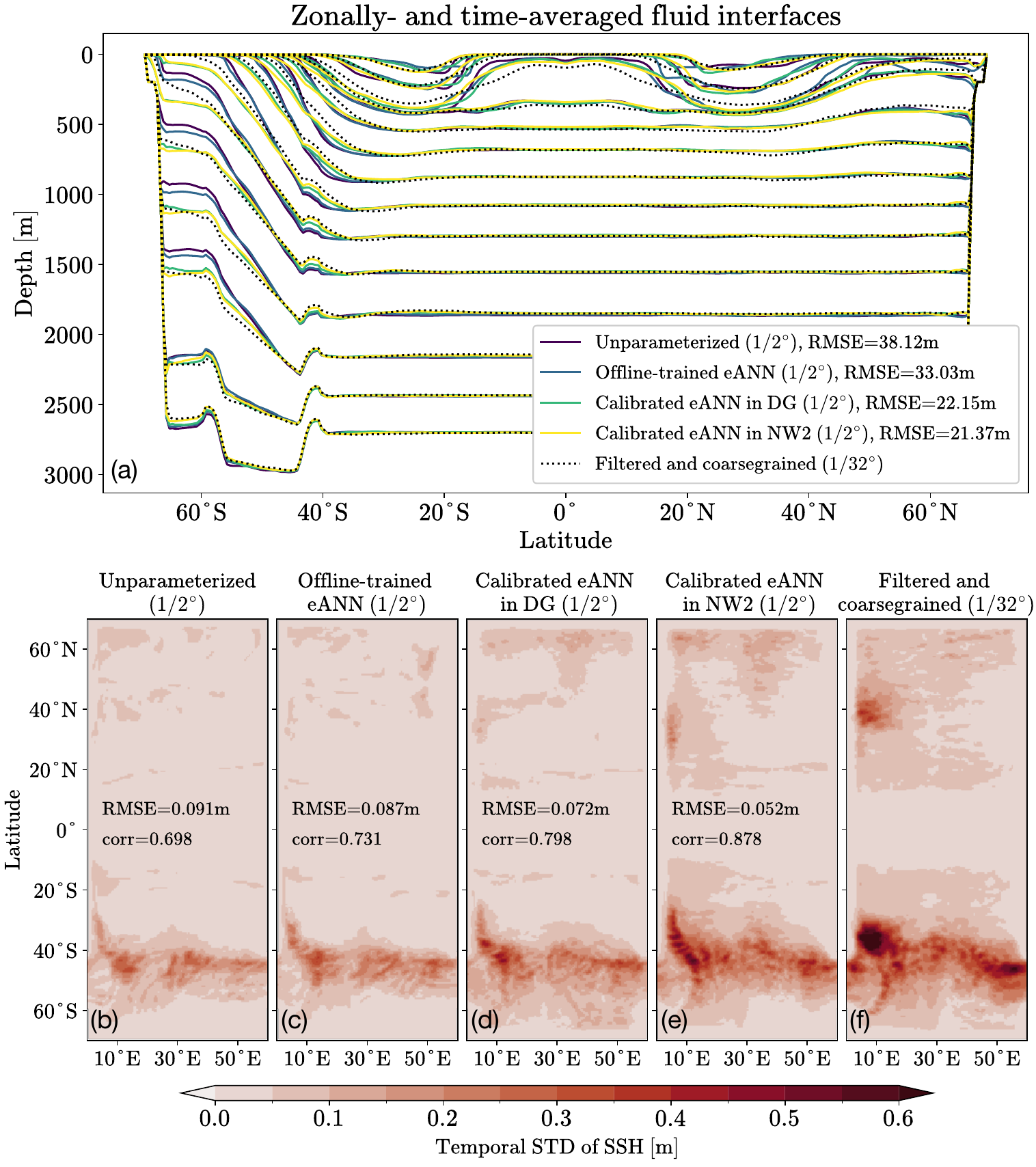}
\caption{Evaluation of calibrated parameterizations in 30000-day simulations in configuration NeverWorld2. (a) Zonally- and time-averaged vertical coordinate of internal fluid interfaces. Lower row shows temporal standard deviation of sea surface height for simulations with: (b) unparameterized model, (c) parameterization trained offline with manually adjusted scaling coefficient ($\gamma=0.5$), (d) parameterization calibrated in Double Gyre with manually adjusted $\gamma=0.6$, (e) parameterization calibrated in short 5-year runs in NeverWorld2 configuration, (f) filtered and coarse-grained high-resolution model.
}
\label{fig:nw2}
\end{figure*}

\subsection{Ocean configuration featuring multiple flow regimes} \label{sec:NW2}

We now consider calibrating the eddy parameterization in the NW2 configuration, which features multiple flow regimes. We apply the calibration algorithm developed in the DG configuration with minimal adjustments. These include model initialization protocol and simulation length (described in Section \ref{sec:method}), adjusting the scheduler step ($\delta t$) and ensemble size ($n_e$) (Section \ref{sec:loss}), and reducing the number of iterations to 4.

Below, we show the evaluation of coarse ocean models, which are integrated to the statistical equilibrium for 30000 days from a state of rest and use the last 800 days to compute statistics. 

The unparameterized coarse ocean model displays strong biases in the mean ocean state, see the position of fluid interfaces in the idealized Southern Ocean (60$^{\circ}$S-40$^{\circ}$S), and too low variability of the sea surface height (Figure \ref{fig:nw2} (a,b)). Offline-trained parameterization allows only for a slight reduction in the mean state bias (RMSE in averaged interfaces drops by $13.4\%$), leaving variability effectively unchanged (RMSE in SSH STD reduces by $4.4\%$), see Figure \ref{fig:nw2} (c). The parameterization calibrated in the DG configuration is more skillful in reducing the model biases: the mean state was improved by $41.9\%$, and the variability was improved by $20.9\%$ (Figure \ref{fig:nw2} (d)). In these two parameterized models, we tuned the scaling coefficient ($\gamma$) to minimize the error on the presented metrics; see Figures S6 and S7 in the SI. That way, we evaluate the effectiveness of the spatial pattern predicted by the parameterization, leaving the coefficient as a tunable parameter that often does not generalize across configurations and must be tuned manually \cite{perezhogin2024stable, balwada2025design, kamm2026assessing}.

We now evaluate the eddy parameterization calibrated in the NW2 configuration in short simulations. This parameterization does not require further adjustment of the scaling coefficient $\gamma$ and can be effectively applied in long 30000-day simulations. The ocean model with this parameterization achieves a similar error in the mean state compared to the parameterization calibrated in a simpler configuration (Figure \ref{fig:nw2}(a)). However, it improves the error and the pattern correlation of the induced variability by $27.8\%$ and $10\%$, respectively (Figure \ref{fig:nw2}(e)). This demonstrates that the calibration algorithm effectively learns the spatial patterns specific to the NW2 configuration.

One of the major advantages of using the calibrated parameterizations is the improvement of the deep Southern Ocean stratification (Figure \ref{fig:nw2} (a)). Some challenges remain. The calibrated parameterization does not sufficiently enhance variability in the western boundary current extension (30$^{\circ}$N–50$^{\circ}$N $\times$ 0$^{\circ}$E–20$^{\circ}$E) (Figure \ref{fig:nw2}(e)) and has limited improvement in the isopycnal structure in the upper 500m of tropical and Southern ocean (Figure \ref{fig:nw2}(a)). This occurs because the deep Southern Ocean dominates the loss function, thereby reducing the weight assigned to other dynamically important regions. The presence of multiple dynamical regimes therefore renders the optimization problem effectively overconstrained. 


%

\section{Discussion}
In this study, we explored the effectiveness of a calibration approach in improving the mean state and variability of an idealized ocean model by adjusting parameters of the data-driven mesoscale eddy parameterization of \citeA{perezhogin2025generalizable}.

We found that the Ensemble Kalman Inversion (EKI) can significantly improve the errors in the mean state and variability at a challenging resolution of $1/2^{\circ}$. The improvement amounts to factors of $2.2-3.3$ in the DG configuration and a factor of $1.7$ in the NW2 configuration compared to the unparameterized model. Further, the improvement amounts to factors of $1.3-2.3$ in the DG configuration and $1.5-1.7$ in the NW2 configuration, compared to the offline-trained parameterization with the optimally tuned coefficient. The parameterization calibrated in a simple configuration (DG) can effectively reduce model biases in an unseen configuration (NW2) after manually adjusting the scaling coefficient. This demonstrates that the spatial patterns learned by the calibrated parameterization generalize well across ocean configurations. Nevertheless, we show that calibrating in the NW2 configuration directly further reduces the variability error by a factor of $1.4$ compared to the parameterization calibrated in a simpler configuration.

The calibration algorithm modifies the offline-trained parameterization as follows (Figure S8 in SI). The predicted along-gradient fluxes exhibit markedly different spatial patterns, with greater skewness towards stronger upgradient fluxes (backscatter). Further, the cross-gradient fluxes were attenuated by $30-40\%$ without changing the spatial pattern, and the isotropic stress remained almost unchanged. A similar effect cannot be achieved by tuning a single coefficient in front of the offline-trained parameterization. The considerable change in the spatial pattern of the along-gradient fluxes can be explained as follows. First, the stronger upgradient fluxes compensate for the excessive numerical dissipation (biharmonic Smagorinsky) that was not observed during training. Second, the along-gradient fluxes, both up- and downgradient, produce strong responses that project onto the ocean model biases \cite{griffies2000biharmonic, fox2008can, jansen2014parameterizing}.

We demonstrated that the EKI method is robust to noise in time-averaged statistics arising from the ocean's chaotic dynamics. Furthermore, the calibration can be performed efficiently in 5-year simulations without integrating to statistical equilibrium (approximately 100 years) if an accurate estimate of the initial ocean state is provided.

The advantages of the EKI algorithm include convergence in a few iterations and parallelism across ensemble members. The computational cost of the calibration in the DG configuration is only 1,000 CPU-core hours, with each simulation requiring 1 CPU core for 1 hour, and 1,000 simulations performed. The small computational cost allowed us to repeat the calibration 50-100 times while experimenting with the calibration protocol, including the loss function, noise model, parameters to be calibrated, and their prior distributions. Once the calibration problem is set up, tuning of the EKI parameters (scheduler step and ensemble size) is straightforward. The computational cost of the calibration in the NW2 configuration is much higher—50,000 CPU-core hours. This highlights the importance of assessing the calibration protocol in a simple, idealized configuration.

We found that efficient calibration can be achieved by perturbing a subset of the neural network's parameters by roughly 25\% on average (Table S2 in the SI). This emphasizes that the offline training plays a significant role in the calibration process, as it determines the weights and biases of the first layer of the neural network and provides initialization for calibrating the last layer.

We found that calibration in a configuration with only one flow regime tends to be underconstrained, whereas calibration in a configuration with multiple flow regimes tends to be overconstrained. In the former case, a careful validation protocol is essential to prevent overfitting \cite{williamson2017tuning}. In the latter case, improvements from calibration are largely concentrated in the ocean's most energetic region. Achieving improvements across multiple flow regimes ultimately requires introducing additional parameters into the calibration procedure, which can be done in various ways \cite{prevost2025detection, tuppi2023simultaneous, abernathey2013global, liu2012adjoint, hallberg2013using}. However, this direction must be pursued with caution, as it may increase the risk of overfitting and reduce generalization capabilities \cite{maddison2024online}.


Our calibration protocol, which bypasses the integration to statistical equilibrium, represents a compromise between optimizing weather forecasting skill \cite{kochkov2021machine,
frezat2022posteriori,
ouala2024online, Kochkov2024, 
maddison2024online} and optimizing the climate metrics in statistical equilibrium \cite{williamson2017tuning, mrozowska2025bayesian,   dunbar2021calibration}. The method can be extended to improve the mean state and variability of realistic ocean models using modern data-assimilation systems \cite{delworth2020spear} or reanalysis products \cite{jean2021copernicus}, which provide estimates of the ocean state. We emphasize that short simulations (a few years) are not suitable for inferring parameters governing processes that evolve over millennial timescales, such as changes in background vertical diffusivity. However, the proposed approach can provide an initial estimate of parameters controlling processes that act on interannual timescales, particularly those influencing the air–sea interface.


Our calibration framework provides a systematic approach for improving parameterizations and reducing biases in global ocean models built on the MOM6 dynamical core, including those used by GFDL \cite{adcroft2019gfdl} and NCAR \cite{grooms2024stochastic}.

\section*{Conflict of Interest}
The authors declare no conflicts of interest relevant to this study.

\section*{Open Research Section}
The calibration script, eANN weights, and plots are available at \citeA{pavel_perezhogin_2026_18809537}. The simulation data and the offline analysis can be found at \citeA{perezhogin_2026_18809615}. Ensemble Kalman Inversion library is available at \citeA{oliver_dunbar_2025_16616326}.

\acknowledgments
This project is supported by Schmidt Sciences, as part of the M$^2$LInES project.
This research was also supported in part through the NYU IT High Performance Computing resources, services, and staff expertise. We would like to acknowledge high-performance computing support from the Derecho system, project code UOSC0036, provided by the NSF National Center for Atmospheric Research (NCAR), sponsored by the National Science Foundation. We are grateful to Dhruv Balwada and Fabrizio Falasca for their valuable comments and suggestions. We also thank the editor, Dr. Gregory Johnson, as well as two anonymous reviewers for their constructive comments that helped improve the quality and presentation of this paper.

\bibliography{agusample}


%
%
%
%
%

\end{document}


\makeatletter
\def\@makecol{\setbox\@outputbox
     \vbox{\boxmaxdepth \maxdepth
\ifdim\ht\dbltopins<1pt\else\unvbox\dbltopins\fi
     \unvbox\@cclv
\ifdim\ht\dblbotins<1pt\else\unvbox\dblbotins\fi%
\ifvoid\footins\else\vskip\skip\footins\footnoterule\unvbox\footins\fi
\vskip 0pt plus 1fil minus \maxdimen
}%
\global\savedblfigandtabnumber\dblfigandtabnumber
   \xdef\@freelist{\@freelist\@midlist}\gdef\@midlist{}\@combinefloats
   \setbox\@outputbox\vbox to\@colht{\boxmaxdepth\maxdepth
   \@texttop\dimen128=\dp\@outputbox\unvbox\@outputbox
   \vskip-\dimen128\@textbottom}%
   \global\maxdepth\@maxdepth}
\makeatother

\title{Supporting Information for "Calibration of a neural network ocean closure for improved mean state and variability"}

\vspace{0.5 cm}

\authors{Pavel Perezhogin\affil{1}, Alistair Adcroft\affil{2}, Laure Zanna\affil{1}}

\affiliation{1}{Courant Institute School of Mathematics, Computing, and Data Science, New York University, New York, NY, USA}

\affiliation{2}{Program in Atmospheric and Oceanic Sciences, Princeton University, Princeton, NJ, USA}

\begin{article}

\noindent\textbf{Contents of this file}
\begin{enumerate}
\item Text S1 to S4
\item Tables S1 to S2
\item Figures S1 to S8
\end{enumerate}



\newpage

\section{*}{Text S1. Implementation of equivariant steerable convolutional neural network (esCNN) in the MOM6 ocean model}
We constrain the neural network to be equivariant to rotations and reflections by imposing these symmetries as hard constraints, using an equivariant steerable convolutional neural network (esCNN) \cite{e2cnn, cesa2022a}, implemented in the software package \texttt{escnn} (version 1.0.11).
The replacement of the neural network used in \citeA{perezhogin2025generalizable} with a steerable CNN  does not require reimplementation of the original parameterization to the MOM6 ocean model. We simply replace the file with weights and biases for the online implementation as follows. An equivariant CNN, which operates with equivariant convolutional filters, can be transformed to a regular CNN, which operates with arbitrary convolutional filters, and this transformation is provided by the \texttt{escnn} package. Furthermore, a regular CNN can be transformed to an ANN when all hidden layers of the CNN have a kernel size $1 \times 1$. This transformation is done by simply reshaping the CNN kernels from 4D tensors to 3D tensors. Implementation of the esCNN without changing the source code of the MOM6 is desirable, as the parameterization of \citeA{perezhogin2025generalizable} is already a part of the MOM6 source code, and thus modeling centers can use our new parameterization by simply replacing the weights and biases file. See Table S1 and Figure S1 for comparison of the architecture and performance of the neural network used in this study and in \citeA{perezhogin2025generalizable}.

\section{*}{Text S2. How equivariance is achieved in esCNN}
An important property of our new parameterization is that it simply represents a more accurate way to choose tunable parameters (weights and biases), as opposed to the original parameterization of \citeA{perezhogin2025generalizable}, but this choice does not change the structure of the machine-learning model.


Consider one convolutional layer with the kernel size $3 \times 3$, which receives one scalar field as an input and returns one scalar field as an output. Examples of scalar fields include pressure, temperature, and the trace of the stress tensor. An example of an arbitrary convolutional layer (not constrained) with randomly generated weights can be given by the $3 \times 3$ matrix as shown below:
\begin{equation}
    \begin{bmatrix}
        0.2083 &  0.1543 & -0.1277 \\
        0.3257 & -0.1831 & 0.0486, \\
        -0.2505 &  0.0157 & -0.2322
    \end{bmatrix}.
\end{equation}

Imposing the equivariance property $D_8$, which is equivariance to the rotations that are multiples of $45^{\circ}$ and reflections, results in the following randomly generated convolutional layer:
\begin{equation}
    \begin{bmatrix}
        0.0230 & 0.0393 & 0.0230 \\
        0.0393 & 0.2398 & 0.0393 \\
        0.0230 & 0.0393 & 0.0230
    \end{bmatrix}.
\end{equation}
As we can see, the constrained convolutional layer has only 3 unique numbers.  
We provide the Python code for how to generate the equivariant convolutional kernel shown above:
\begin{Verbatim}[fontsize=\small, frame=single]
from escnn import gspaces, nn
import torch
torch.manual_seed(42)
r2_act = gspaces.flipRot2dOnR2(N=8)
feat_type_in = nn.FieldType(r2_act, [r2_act.trivial_repr])
feat_type_out = nn.FieldType(r2_act, [r2_act.trivial_repr])
layer = nn.R2Conv(feat_type_in, feat_type_out, kernel_size=3)
print(layer.expand_parameters())
\end{Verbatim}

In our study, we calibrate only free parameters in an equivariant kernel, instead of all parameters (in this example, 9). A set of free parameters is provided by the \texttt{escnn} package 
\begin{Verbatim}[fontsize=\small, frame=single]
for p in layer.parameters():
    print(p)
\end{Verbatim}
\noindent simplified output:
\begin{Verbatim}[fontsize=\small, frame=single]
tensor([0.2381, 0.0911]).
\end{Verbatim}

Thus, implementing equivariances considerably reduces the number of parameters to be calibrated (in this example, from 9 to 2, i.e., by 77\%). See Table S1 for the reduction in the number of parameters in the eANN parameterization used in this study.

\section{*}{Text S3. Choice of a group of equivariances for the neural network parameterization}
Our goal is to constrain the neural network as much as possible, such that unphysical predictions would not appear as a result of the minimization of the online loss function. Additionally, using a bigger group of equivariances reduces the number of free parameters for a fixed size of the neural network (see Text S2), which simplifies the calibration problem and improves generalization.

A popular choice of a group of equivarainces implemented for the subgrid parameterizations is $C_4$, that is, multiples of rotations by $90^{\circ}$ \cite{guan2022learning, pawar2023frame, connolly2025deep}. Here, we additionally include reflections (same as \citeA{perezhogin2025generalizable}), and extend the rotation group to multiples of $45^{\circ}$, i.e., group $D_8$ consisting of 16 transformations. Our choice is explain below.

Because our parameterization maps the velocity gradient tensor to the stress tensor, we need to take additional care, as these are not scalar fields. We need to organize the information stored in these tensors into groups of one field or a pair of fields (so-called irreducible representations). Examples of irreducible representations are scalar field (trace of the tensor), pseudoscalar (vorticity, which is invariant to rotation but changes sign under reflection), and vector field (velocity). There are additional irreducible representations that require further explanation with examples.

\paragraph{Transformation of the tensor components under coordinate transformation} 
We consider the $2\times 2$ stress tensor (note the analysis below applies to any tensor):
\begin{equation}
    \mathbf{T} = T_{xx} \mathbf{e}_{x} \otimes \mathbf{e}_x + 
    T_{yy} \mathbf{e}_{y} \otimes \mathbf{e}_y + T_{xy} \mathbf{e}_{x} \otimes \mathbf{e}_y + T_{yx} \mathbf{e}_{y} \otimes \mathbf{e}_x,
\end{equation}
here $T_{xx}$, $T_{yy}$, $T_{xy}$, $T_{yx}$ are 4 components of the tensor, and $\mathbf{e}_x$ and $\mathbf{e}_y$ are Cartesian coordinate basis vectors.

We need to understand how the components of an arbitrary tensor transform under the rotation of the coordinate system by $45^{\circ}$:
\begin{gather}
    \mathbf{e}_x = \frac{1}{\sqrt{2}} \left( \mathbf{e}_x' - \mathbf{e}_y' \right). \\
    \mathbf{e}_y = \frac{1}{\sqrt{2}} \left( \mathbf{e}_x' + \mathbf{e}_y' \right),
\end{gather}
where $\mathbf{e}_x'$ and $\mathbf{e}_y'$ represent basis vectors of a new Cartesian coordinate system, see illustration below.

{\centering
\includegraphics[scale=0.5]{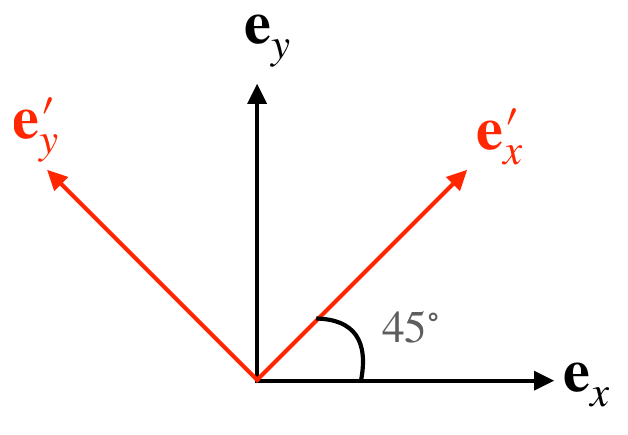}
\par}

\noindent The tensor $\mathbf{T}$ introduced above in a new coordinate system reads as:
\begin{gather}
    2 \mathbf{T} = \\
    T_{xx} \left( \mathbf{e}_x' - \mathbf{e}_y' \right) \otimes \left( \mathbf{e}_x' - \mathbf{e}_y' \right) + \\ 
    T_{yy} \left( \mathbf{e}_x' + \mathbf{e}_y' \right) \otimes\left( \mathbf{e}_x' + \mathbf{e}_y' \right) + \\T_{xy} \left( \mathbf{e}_x' - \mathbf{e}_y' \right) \otimes \left( \mathbf{e}_x' + \mathbf{e}_y' \right) + \\
    T_{yx} \left( \mathbf{e}_x' + \mathbf{e}_y' \right) \otimes \left( \mathbf{e}_x' - \mathbf{e}_y' \right) = \\
    (T_{xx} + T_{yy} + T_{xy} + T_{yx}) \mathbf{e}_x' \otimes \mathbf{e}_x' + \\
    (T_{xx} + T_{yy} - T_{xy} - T_{yx}) \mathbf{e}_y' \otimes \mathbf{e}_y' + \\
    (-T_{xx} + T_{yy} + T_{xy} - T_{yx}) \mathbf{e}_x' \otimes \mathbf{e}_y' + \\
    (-T_{xx} + T_{yy} - T_{xy} + T_{yx}) \mathbf{e}_y' \otimes \mathbf{e}_x'
\end{gather}

\noindent Now let's introduce the tensor components in the new coordinate system:
\begin{equation}
    \mathbf{T} = T_{xx}' \mathbf{e}_{x}' \otimes \mathbf{e}_x' + 
    T_{yy}' \mathbf{e}_{y}' \otimes \mathbf{e}_y' + T_{xy}' \mathbf{e}_{x}' \otimes \mathbf{e}_y' + T_{yx}' \mathbf{e}_{y}' \otimes \mathbf{e}_x'.
\end{equation}

\noindent Then, the transformation of tensor components reads as:
\begin{gather}
    T_{xx'} = \frac{1}{2} (T_{xx} + T_{yy} + T_{xy} + T_{yx}), \\
    T_{yy'} = \frac{1}{2} (T_{xx} + T_{yy} - T_{xy} - T_{yx}), \\
    T_{xy'} = \frac{1}{2}  (-T_{xx} + T_{yy} + T_{xy} - T_{yx}), \\
    T_{yx'} = \frac{1}{2} (-T_{xx} + T_{yy} - T_{xy} + T_{yx}).
\end{gather}
\noindent or in matrix form
\begin{equation}
    \begin{bmatrix}
        T_{xx'}\\
        T_{yy'}\\
        T_{xy'}\\
        T_{yx'}\\
    \end{bmatrix} = 
    \left(\frac{1}{2}
    \begin{bmatrix}
        1 & 1 & 1 & 1 \\
        1 & 1 & -1 & -1 \\
        -1 & 1 & 1 & -1 \\
        -1 & 1& -1 & 1 
    \end{bmatrix} \right)
    \begin{bmatrix}
        T_{xx}\\
        T_{yy}\\
        T_{xy}\\
        T_{yx}\\
    \end{bmatrix}. \label{eq:transformation}
\end{equation}

\paragraph{Irreducible representations}
Eigenvectors of the transformation (Eq. \eqref{eq:transformation}) are the irreducible representations. We found that there are two fields that transform as scalar fields (eigenvalues equal 1):
\begin{gather}
    T_T = \frac{1}{2} \left(T_{xx} + T_{yy} \right), \\
    T_{\Omega} = \frac{1}{2} \left(T_{xy} - T_{yx} \right)
\end{gather}
In a case when the tensor $\mathbf{T}$ is the velocity gradient tensor, these irreducible representations are horizontal divergence and horizontal vorticity, respectively. In a case when $\mathbf{T}$ is the stress tensor, the irreducible representation $T_T$ is the trace of the stress tensor, while $T_{\Omega}$ is always zero for angular momentum conservation. Two additional eigenvectors of Eq. \eqref{eq:transformation} correspond to a pair of complex-conjugate eigenvalues:
\begin{gather}
    T_D = \frac{1}{2} \left(T_{xx} - T_{yy} \right), \\
    T_S = \frac{1}{2} \left(T_{xy} + T_{yx} \right).
\end{gather}
These two components of the velocity gradient tensor and stress tensor are the only two components considered in eddy viscosity parameterizations \cite{griffies2000biharmonic}, see also Eq. 40 in \citeA{bachman2024eigenvalue}. These two components of the tensor must be considered together because they transform as a vector field under coordinate rotation, however, with doubled frequency.

\paragraph{Summary on irreducible representations}
The tensor in a basis of irreducible representations is given below:
\begin{equation}
    \mathbf{T} = \begin{bmatrix}
        T_{xx} & T_{xy} \\
        T_{yx} & T_{yy}
    \end{bmatrix} = 
    \begin{bmatrix}
        T_D + T_T & T_S + T_{\Omega} \\
        T_S - T_{\Omega} & -T_D + T_T
    \end{bmatrix}.
\end{equation}

To sum up, counterclockwise rotation of the coordinate system by $45^{\circ}$ has the following form in a basis of irreducible representations:
\begin{gather}
    T_{T}' = T_{T}, \\
    T_{\Omega}' = T_{\Omega}, \\
    T_D' = T_S, \\
    T_S' = -T_D.
\end{gather}

Similarly, these irreducible representations flip sign under the reflection ($\mathbf{e}_x'=-\mathbf{e}_x$, $\mathbf{e}_y'=\mathbf{e}_y$) as follows:
\begin{gather}
    T_{T}' = T_{T}, \\
    T_{\Omega}' = - T_{\Omega}, \\
    T_D' = T_D, \\
    T_S' = -T_S.
\end{gather}

\paragraph{Choice of symmetry group}
The major reasoning behind our choice of $D_8$ symmetry group is the following. While the physics of geophysical fluids is not rotationally invariant (there is a preferred zonal direction on a $\beta$-plane), and is not reflectionally invariant (there is a preferred vertical direction), we do not pass zonal and vertical directions as input vectors to the parameterization. Thus, the information we passed to the input of the neural network is not enough to infer horizontal and vertical anisotropies of subgrid closure. 
Our way of specifying invariances follows purely from the analysis of the coordinate system, but not from the analysis of the physics of the geophysical flows. This justifies the use of a symmetry group that includes both rotations and reflections. Learning of anisotropic subgrid closures can be enabled by preserving the same symmetry group and including the preferred direction \cite{smith2003anisotropic}. However, this would require changing the structure of the \citeA{perezhogin2025generalizable} parameterization, which we avoid here.

\paragraph{Examples}
Below we show that using the group $D_8$ instead of the group $D_4$ (multiples of rotations by $90^{\circ}$) is essential to eliminate anisotropic closures. $D_4$ allows anisotropic closures because it is a discrete approximation of a continuous group of rotations, and it is less restrictive. Let's consider, as an example, an anisotropic eddy viscosity closure \cite{bachman2024eigenvalue}:
\begin{gather}
    T_D = \sigma_D, \\
    T_S = 0 \cdot \sigma_S.
\end{gather}
Under rotation by $90^{\circ}$, inputs and outputs transform as follows (shown here without derivation):
\begin{gather}
    (-T_D') = (-\sigma_D'), \\
    (-T_S') = 0 \cdot (-\sigma_S').
\end{gather}
Thus, the functional relationship between irreducible representations did not change under this transformation, and $D_4$ group does not prohibit anisotropic closures. Instead, under rotation by $45^{\circ}$ the closure transforms differently:
\begin{gather}
    (-T_S') = (-\sigma_S'), \\
    T_D' = 0 \cdot \sigma_D',
\end{gather}
which has a different functional form. Thus, $D_8$ prohibits anisotropic closures. Slightly anisotropic subgrid closures are often found in data-driven parameterizations; however, human-made decision is often made to eliminate anisotropies if not desired, see \citeA{zanna2020data, ross2022benchmarking}.

Additionally, reflectional invariance in group $D_8$ allows us to prohibit closures that relate vorticity (pseudoscalar) to the trace of the stress tensor (scalar). For example,
\begin{equation}
    T_T = \omega.
\end{equation}
To the best of our knowledge, there are no well-known subgrid closures of this type.

\paragraph{Architecture of the equivariant neural network}
We consider the parameterization that maps three components of the velocity gradient tensor to the three components of the stress tensor:
\begin{equation}
    \sigma_D, \sigma_S, \omega \rightarrow  T_D, T_S, T_T.
\end{equation}
In a language of \texttt{escnn} package, we specify $D_8$ symmetry group as 
\begin{Verbatim}[fontsize=\small, frame=single]
r2_act = gspaces.flipRot2dOnR2(N=8)
\end{Verbatim}
Further, input and output features are assigned the following irreducible representations based on the analysis above:
\begin{itemize}
    \item A pair of ${\sigma}_D = \partial_x {u} - \partial_y {v}$ and ${\sigma}_S = \partial_y {u} + \partial_x {v}$ is \texttt{r2\_act.irrep(1,2)}
    \item $\omega = \partial_x {v} - \partial_y {u}$ is \texttt{r2\_act.irrep(1,0)}
    \item A pair of $T_D = \frac{1}{2} \left(T_{xx} - T_{yy} \right)$ and $T_S = \frac{1}{2} \left(T_{xy} + T_{yx} \right)$ is \texttt{r2\_act.irrep(1,2)}
    \item $T_T = \frac{1}{2} \left(T_{xx} + T_{yy} \right)$ is \texttt{r2\_act.irrep(0,0)}
\end{itemize}
Here, \texttt{r2\_act.irrep(0,0)} is a true scalar, \texttt{r2\_act.irrep(1,0)} is a pseudoscalar, where the first number reads as follows: $0$ means that the sign does not flip under the reflection, and $1$ means that the sign flips under the reflection. The second number (0) is the frequency under coordinate rotation (i.e., these components are invariant under the rotation). Further, \texttt{r2\_act.irrep(1,2)} means that it is a vector which rotates with a double frequency (see number $2$) under coordinate rotation, and one component flips sign under reflection. The hidden layer has regular irreducible representation \texttt{r2\_act.regular\_repr}. The equivariant neural network on a spatial stencil of $3 \times 3$ is defined as:
\begin{Verbatim}[fontsize=\small, frame=single]
from escnn import gspaces, nn
# Set parameters
stencil_size = 3
hidden_layer_size = 64
# Define equivariance group
r2_act = gspaces.flipRot2dOnR2(N=8)
# Inputs: [sigma_D, sigma_S, omega]
feat_type_in  = nn.FieldType(
    r2_act,
    [r2_act.irrep(1,2)] + [r2_act.irrep(1,0)]
)
# Outputs: [T_D, T_S, T_T]
feat_type_out = nn.FieldType(
    r2_act,
    [r2_act.irrep(1,2)] + [r2_act.irrep(0,0)]
)
# Number of transfromations in the group
group_size = r2_act.regular_repr.size
# Hidden layer
feat_type_hid = nn.FieldType(
    r2_act,
    hidden_layer_size//group_size * [r2_act.regular_repr]
)
# esCNN
model = nn.SequentialModule(
    nn.R2Conv(feat_type_in, feat_type_hid, kernel_size=stencil_size),
    nn.ReLU(feat_type_hid),
    nn.R2Conv(feat_type_hid, feat_type_out, kernel_size=1)
)
\end{Verbatim}

\paragraph{Testing}
We experimentally verified the correctness of our equivariant neural network implementation by confirming that it can efficiently learn well-known isotropic closures (e.g., isotropic eddy viscosity; \citeA{zanna2020data}), while failing to learn anisotropic closures or scalar–pseudoscalar relationships.

\section{*}{Text S4. Normalization of fluid interfaces}
The goal of our normalization procedure is to make sure that anomalies of the fluid interfaces are comparable in magnitude. We avoid normalization of the free surface ($\eta_{1/2}$) as it is already on the order of 1. 

Let's consider a two-layer fluid. The quasi-geostrophic stream function in two fluid layers is given by the interface anomalies \cite<see their Eq. 5.89, >{vallis2017atmospheric}:
\begin{gather}
    \eta_{1/2}' = \frac{f}{g} \psi_1, \\
    \eta_{3/2}' = \frac{f}{g_{3/2}'} (\psi_2 - \psi_1),
\end{gather}
where $f$ is the Coriolis frequency, $g$ is the gravity constant and $g_{3/2}'$ is the reduced gravity,  $\psi_1$ is the streamfunction in the upper fluid layer, and $\psi_2$ is the streamfunction in the lower fluid layer. The key observation is that the lower fluid layer is not affected by the wind forcing and is affected by the bottom friction. As a result, the bottom layer is often quiescent ($|\psi_2| \ll |\psi_1|$, \citeA{vallis2017atmospheric}). Thus, eliminating streamfunction from the equations above, we come up with a natural scaling of anomalies of the fluid interfaces:
\begin{equation}
    \eta_{1/2}' \approx - \frac{g_{3/2}'}{g} \eta_{3/2}'.
\end{equation}
We use scaling $\frac{g_{3/2}'}{g}$ for the internal interface of the two-layer fluid.

A similar idea is suggested for a multi-layer fluid, where a quasigeostrophic streamfunction reads as:
\begin{gather}
    \eta_{1/2}' = \frac{f}{g} \psi_1, \\
    \eta_{k+1/2}' = \frac{f}{g_{k+1/2}'} (\psi_{k+1} - \psi_k).
\end{gather}
We again assume that the stream function in the bottom layer is negligibly small ($\psi_{n_z} \approx 0$). Then, we readily find that
\begin{equation}
    \sum_{k=1}^{n_z-1} \frac{g_{k+1/2}'}{g} \eta_{k+1/2}' = \frac{f}{g} \sum_{k=1}^{n_z-1} (\psi_{k+1} - \psi_k) \approx - \frac{f}{g} \psi_1 = - \eta_{1/2}'. \label{eq:interfaces_equality}
\end{equation}
We assume that each term in the sum in Eq. \eqref{eq:interfaces_equality} contributes to the sum equally. Hence, the typical magnitude of a single term ($\frac{g_{k+1/2}'}{g} \eta_{k+1/2}'$) is smaller than that of the free surface by a factor $n_z-1$. We therefore  further multiply each term by $n_z-1$ to make the magnitude of the normalized internal interfaces comparable to that of the free surface:
\begin{equation}
        (n_z-1) \frac{g_{k+1/2}'}{g} \eta_{k+1/2}' \sim  - \eta_{1/2}'.
\end{equation}
This normalization contains the normalization found for a two-layer fluid above as a special case because in that case $n_z=2$.

\end{article}
\newpage

\clearpage

\begin{figure}[h!]
\centering{\includegraphics[width=0.95\textwidth]{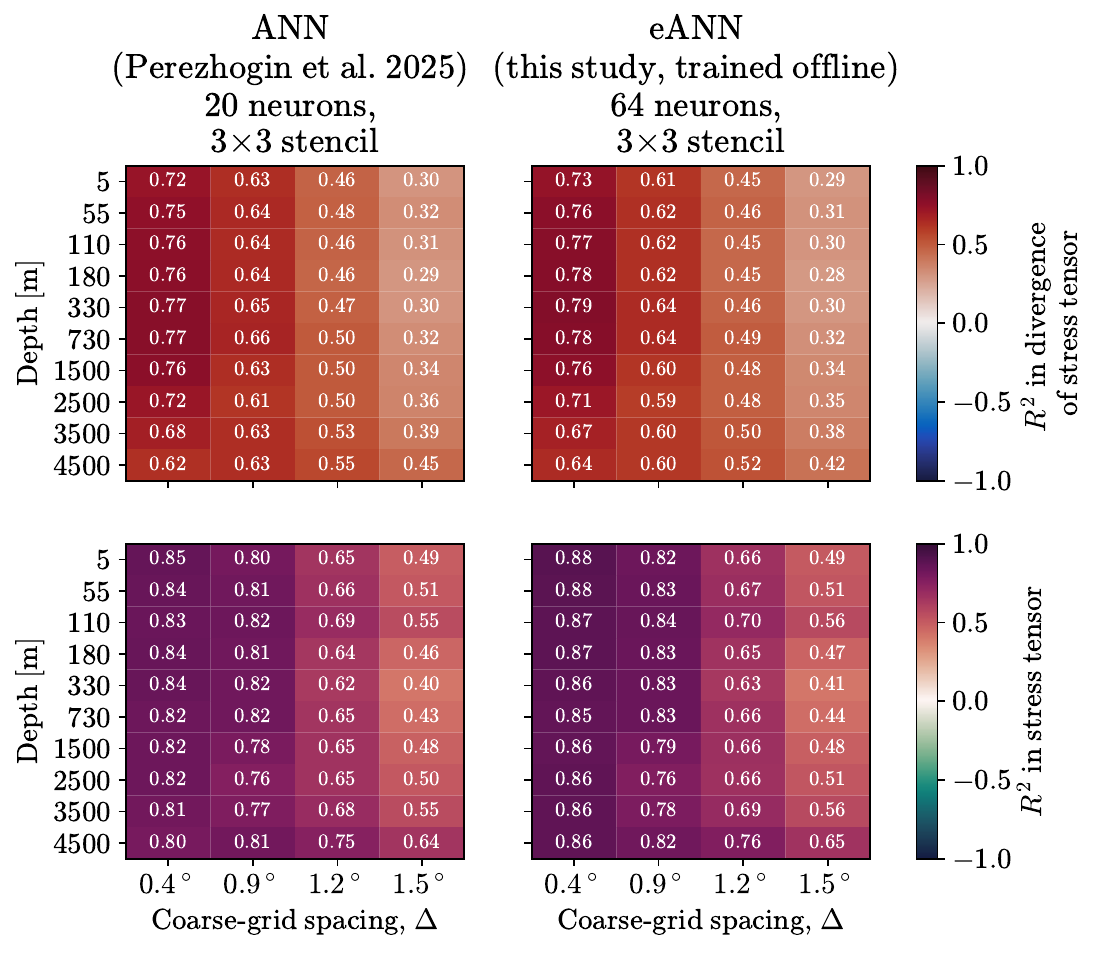}}
\caption{Comparison of the offline skill in the global ocean CM2.6 dataset \cite{griffies2015impacts} between the neural network used for online simulations in \citeA{perezhogin2025generalizable} (on the left) and the equivariant neural network used in this study (on the right). We slightly increased the size of the single hidden layer of the neural network from 20 neurons to 64 neurons and slightly modified the loss function to match the accuracy of predictions in both the divergence of stress tensor ($\nabla \cdot \mathbf{T}$, upper row) and the stress tensor ($\mathbf{T}$, lower row).
}
\end{figure}

\begin{figure}[h!]
\centering{\includegraphics[width=1\textwidth]{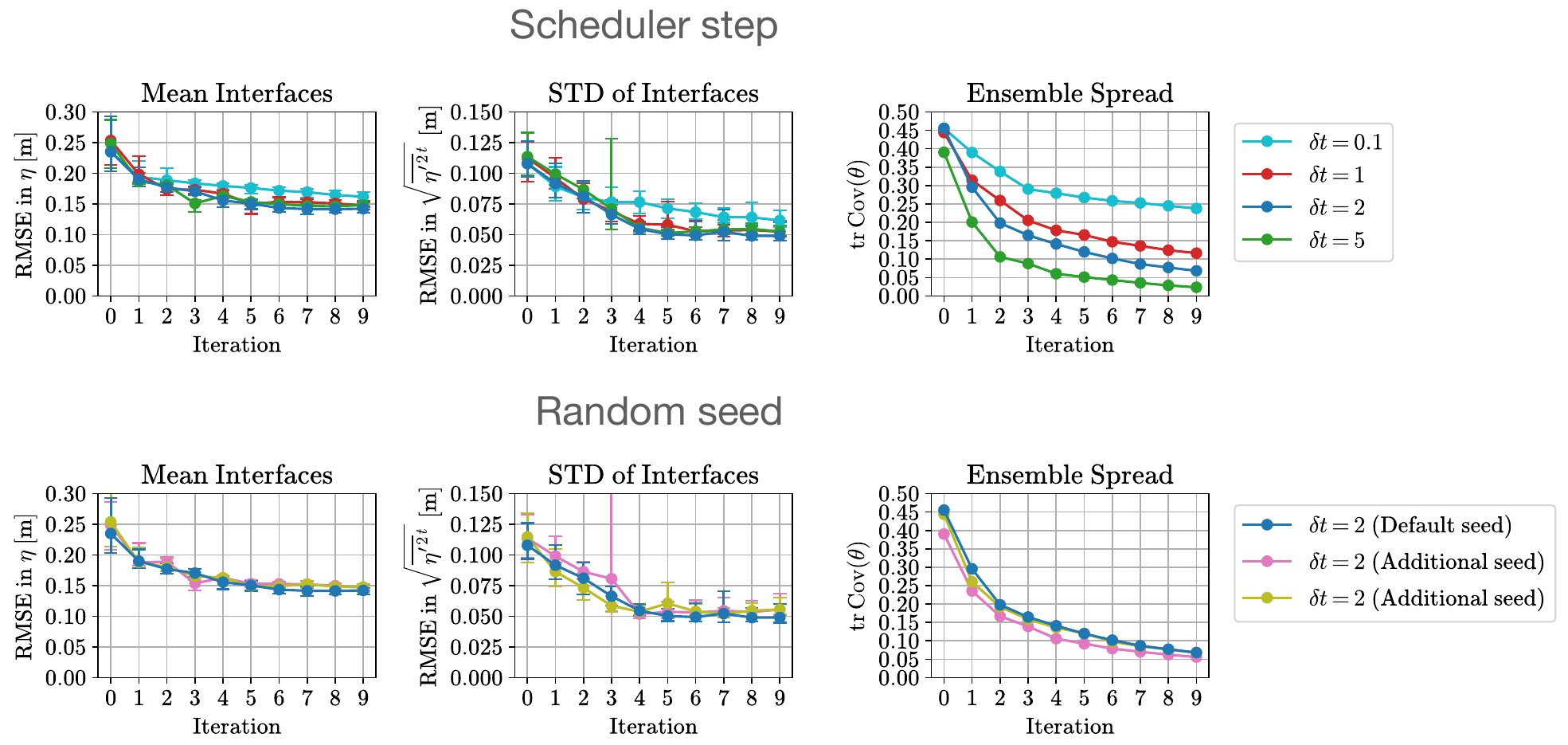}}
\caption{Sensitivity of the calibration in DG configuration to the scheduler step ($\delta t$), in the upper row, and random seed, in the lower row. The left and center columns show the components of the loss function, which track the convergence of the calibration algorithm. The right column shows the spread of the ensemble in parameter space, which tracks the ensemble consensus or collapse. The scheduler step controls the convergence: large values result in a non-monotonic decreasing of the loss function and too fast ensemble collapse. On the contrary, a small value of $\delta t$ results in slower and more monotonic loss decreasing and slower ensemble collapse. Lower panels show that loss dynamics and ensemble spread follow similar trends for two additional random seeds. The ensemble size is $n_e=100$.
}
\end{figure}

\begin{figure}[h!]
\centering{\includegraphics[width=0.85\textwidth]{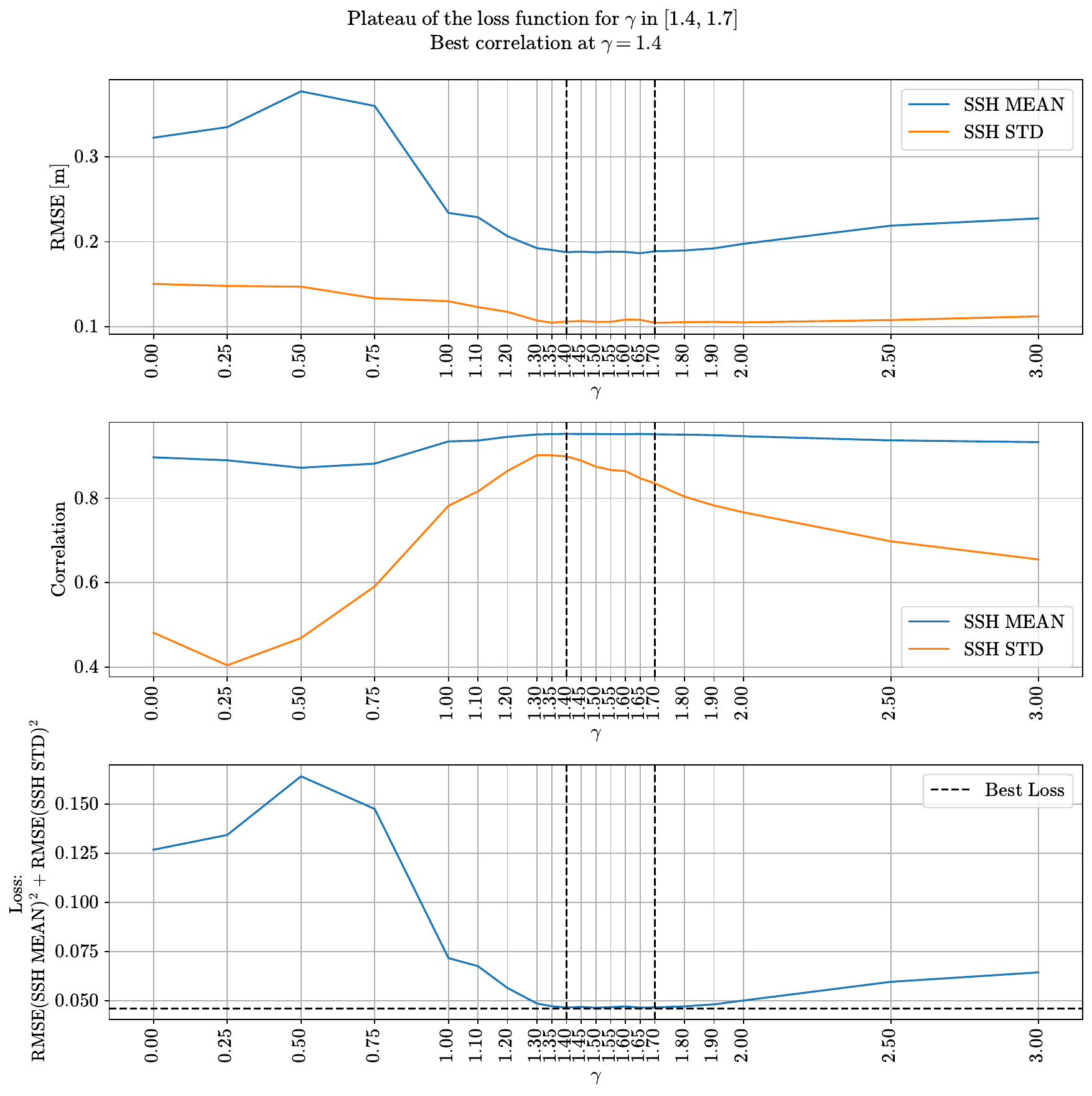}}
\caption{
Calibration of the scaling coefficient $\gamma$ in front of the offline-trained eANN parameterization via a grid search in the Double Gyre configuration in 100-year simulations. Upper panel: root mean squared error (RMSE) in the time-mean sea surface height (SSH MEAN) and temporal standard deviation of sea surface height (SSH STD) between coarse parameterized model and filtered and coarse-grained simulation. Middle panel: spatial pattern correlation. Lower panel: The loss function similar to that used for automatic calibration. Coefficient $\gamma=1.4$ gives the best joint error in the mean state and variability of the surface ocean, and best spatial pattern correlation.
}
\end{figure}

\begin{figure}[h!]
\centering{\includegraphics[width=0.95\textwidth]{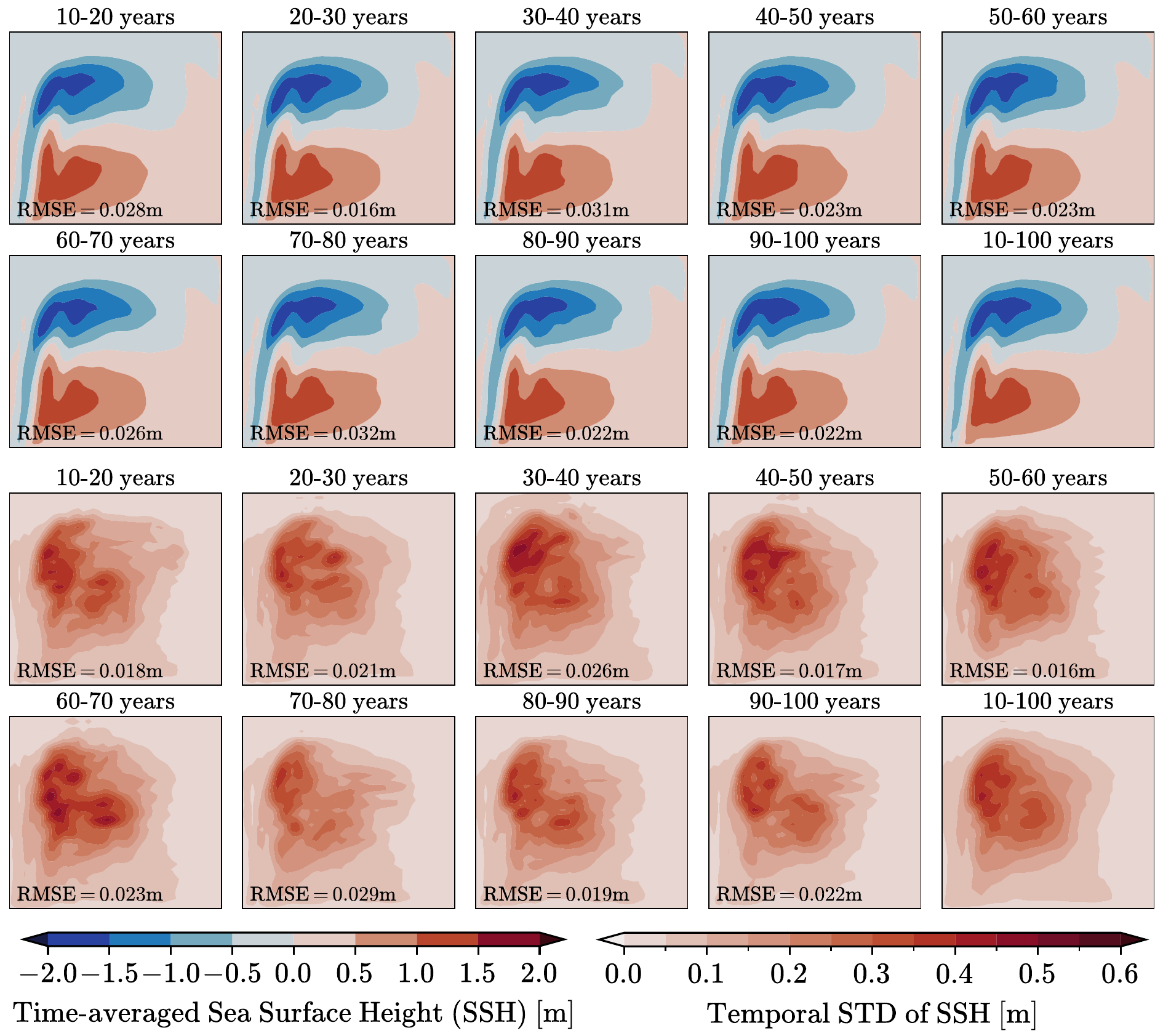}}
\caption{Assessment of the random noise originating from ocean chaotic dynamics in temporal averages. Results for the calibrated eANN parameterization are shown. Panels compare 10-year averages over non-overlapping time intervals to the 90-year average. Root Mean Squared Error (RMSE) is shown with respect to the 90-year average. The noise magnitudes in the temporal average and standard deviation (STD) of the sea surface height (SSH) are comparable ($\mathrm{RMSE}\approx 0.02$). Note that the noise is more significant for the SSH STD metric, as it amounts to approximately half of the parameterized model error. Noise in the time-averaged SSH looks less significant by visual inspection. However, we note that the noise is hidden by the strong mean pattern.
}
\end{figure}

\begin{figure}[h!]
\centering{\includegraphics[width=0.95\textwidth]{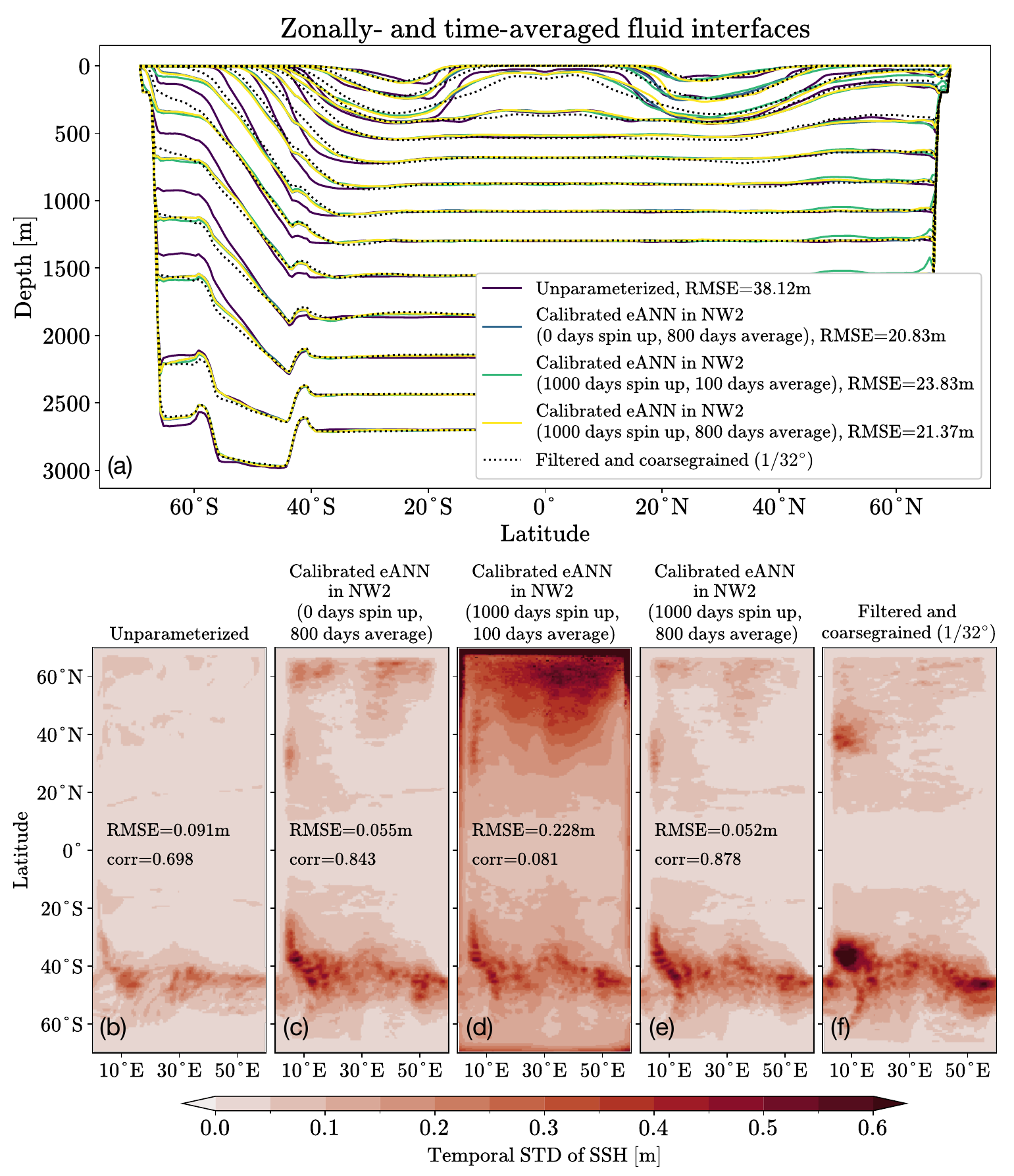}}
\caption{
Sensitivity to the setup of the calibration protocol in NW2 configuration: spin-up time (0 or 1000 days) and averaging time (800 or 100 days). The metrics shown are obtained in long 30000-days integrations from the state of rest. Increasing the averaging time is crucial for improved variability (d,e). The spin-up time has almost no effect on both the mean state and variability (c,e).
}
\end{figure}

\begin{figure}[h!]
\centering{\includegraphics[width=0.95\textwidth]{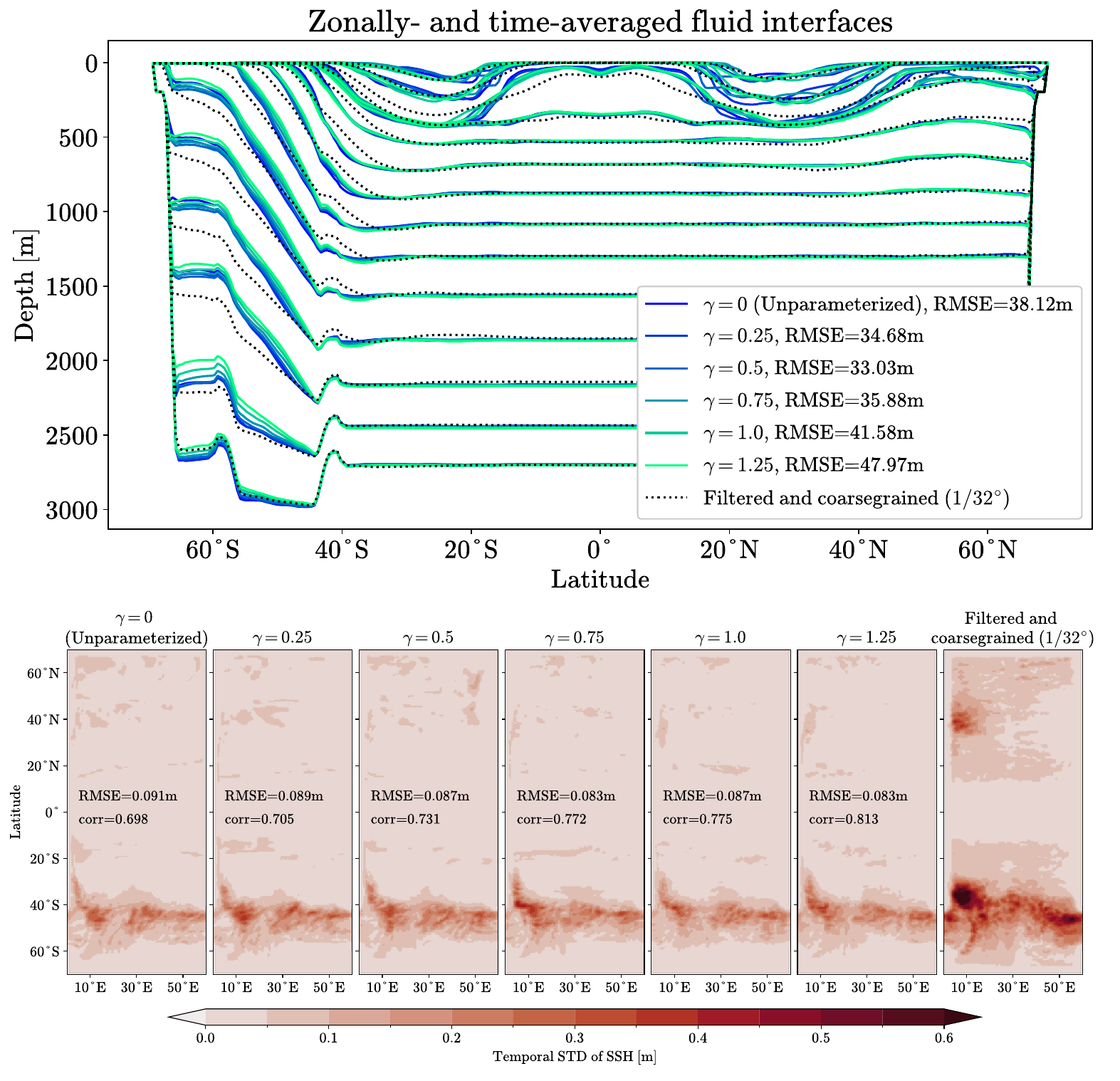}}
\caption{Mean state and variability in NW2 configuration at resolution $1/2^{\circ}$ for the offline-trained eANN parameterization. Sensitivity to the scaling coefficient ($\gamma$) in front of the parameterization, where $\gamma=0$ corresponds to the unparameterized model. The mean state and variability can be improved by $13.4\%$ (upper row) and $4.4\%$ (lower row), respectively, compared to the unparameterized simulation when $\gamma=0.5$.
}
\end{figure}

\begin{figure}[h!]
\centering{\includegraphics[width=0.95\textwidth]{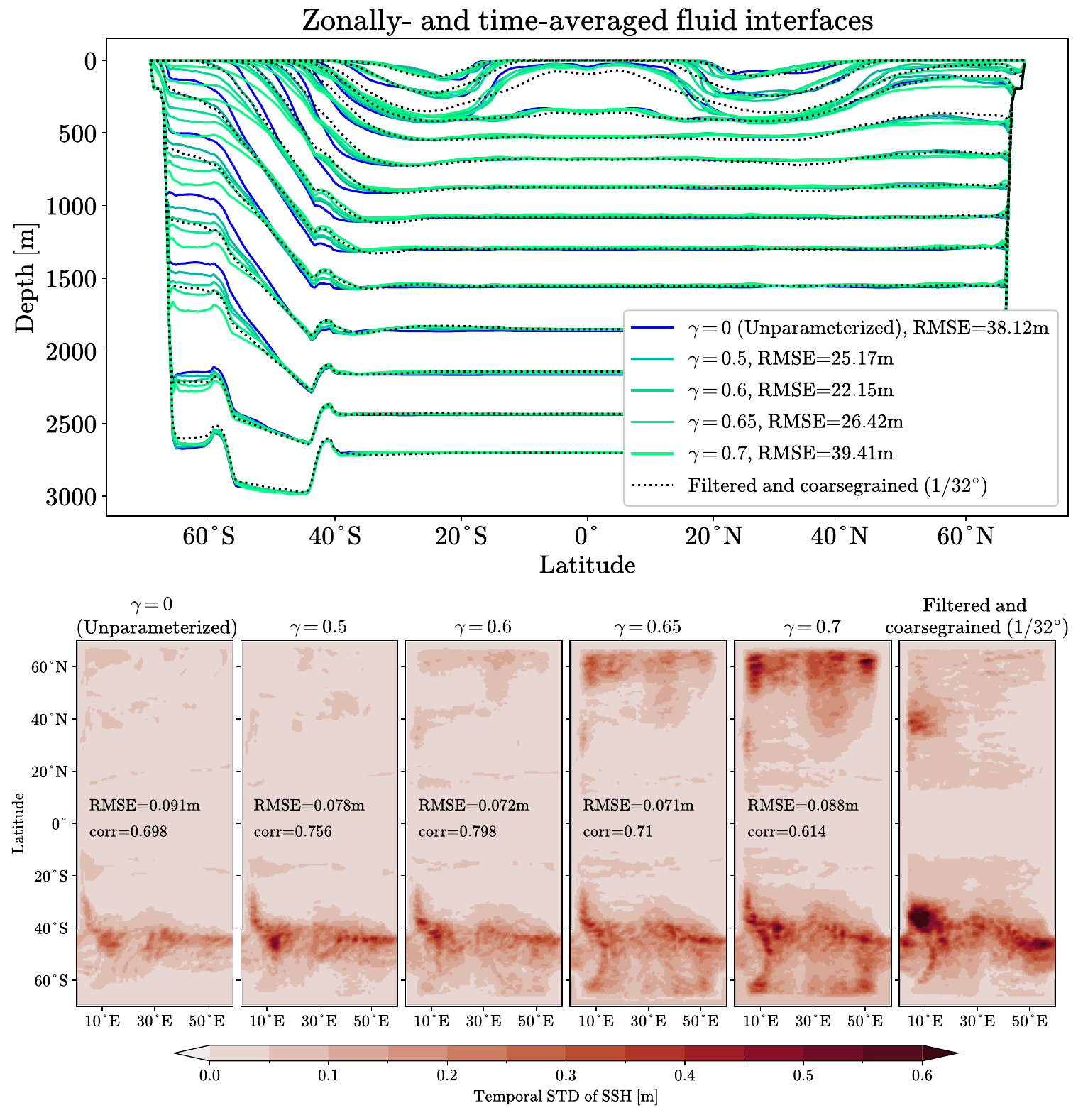}}
\caption{Generalization to unseen configuration NW2 for the eANN parameterization calibrated in different configuration (DG). Sensitivity to the scaling coefficient ($\gamma$) in front of the parameterization, where $\gamma=0$ corresponds to the unparameterized model. The mean state and variability can be improved by $41.9\%$ (upper row) and $20.9\%$ (lower row), respectively, compared to the unparameterized simulation when $\gamma=0.6$.
}
\end{figure}

\begin{figure}[h!]
\centering{\includegraphics[width=0.99\textwidth]{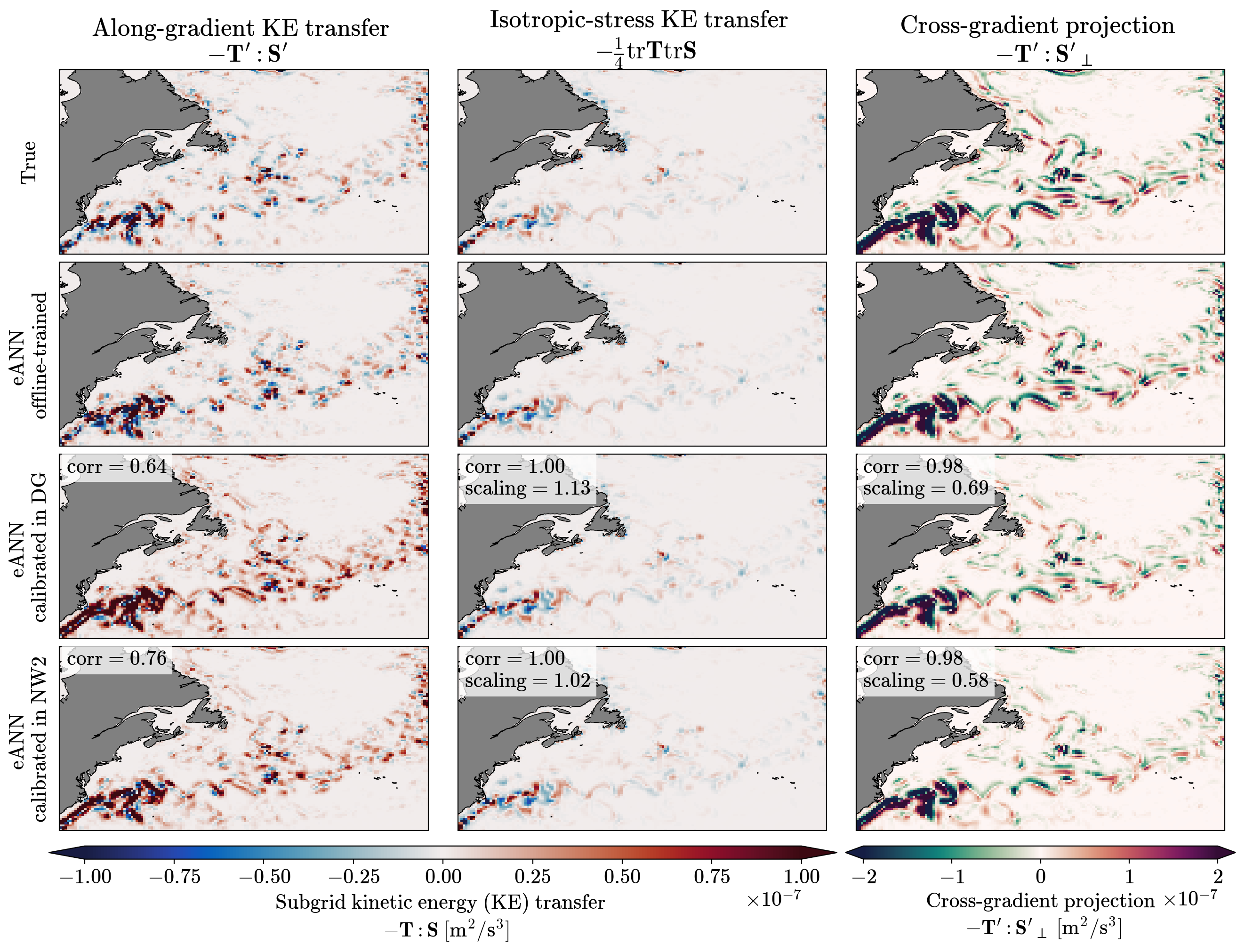}}
\caption{
Offline analysis of parameterizations on output from the CM2.6 climate model at coarse-graining factor 4 and depth 5 m. $\mathbf{S}=\frac{1}{2}(\nabla \mathbf{u}+(\nabla \mathbf{u})^T)$ is the strain-rate tensor, and $\mathbf{u}\partial_t\mathbf{u}=\cdots-\mathbf{T}:\mathbf{S}$ is the subgrid KE transfer, where positive (red) values indicate backscatter and negative (blue) values are dissipation. The KE transfer is decomposed into the contributions from the along-gradient fluxes (left) and isotropic stress (center), where $\mathbf{T}'=\mathbf{T}-\frac{1}{2}\mathrm{tr}\mathbf{T}$ is the deviatoric stress. The cross-gradient projection (right) uses $\mathbf{S}'_{\perp}=\frac{1}{2}\begin{bmatrix}\sigma_S & -\sigma_D \\ -\sigma_D & -\sigma_S\end{bmatrix}$ which satisfies $\mathbf{S}'{\perp} :\mathbf{S}'=0$. Rows show KE fluxes from the high-resolution data (top), the offline-trained eANN with $\gamma=1$ (second), and eANNs calibrated in DG and NW2 configurations (bottom two). The “corr” denotes pattern correlation with the offline-trained predictions. The scaling coefficient shows how to rescale the offline-trained parameterization to obtain the calibrated one.
}
\end{figure}

\clearpage

\begin{table}[]
\begin{tabular}{|l|c|c|}
\hline                                     & Perezhogin et al. 2025 & eANN (this study) \\ \hline
Input features                            & 27                     & 27                \\ \hline
Output features                           & 3                      & 3                 \\ \hline
Hidden layers                             & 1                      & 1                 \\ \hline
Neurons in hidden layer                   & 20                     & 64                \\ \hline
Parameters in the first layer             & 560                    & 1792              \\ \hline
Parameters in the last layer              & 63                     & 195               \\ \hline
Equivariant parameters in the first layer & N/A                    & 76                \\ \hline
Equivariant parameters in the last layer  & N/A                    & 13                \\ \hline
\end{tabular}
\caption{Comparison of the architecture of the neural networks used in \citeA{perezhogin2025generalizable} for online simulations and the equivariant neural network (eANN) used in this study.}
\end{table}

\begin{table}[]
\begin{tabular}{|l|cccccccccccc|c|c|}
\hline
Parameter & \multicolumn{12}{c|}{Weights} & Biases & $\gamma$ \\ \hline

Offline
& 0.68 & 0.69 & 1.15 & 0.15 & 0.60 & -0.60
& 0.74 & 0.01 & -0.36 & 0.31 & -0.04 & 2.25
& 0.15 & 1.00 \\ \hline

STD
& \multicolumn{12}{c|}{0.18} 
& 0.04 & 0.25 \\ \hline

DG
& 0.64 & 0.66 & 1.35 & 0.23 & 0.15 & -0.50
& 0.40 & 0.29 & -0.79 & 0.51 & -0.19 & 2.06
& 0.16 & 1.08 \\ \hline

NW2
& 0.85 & 0.73 & 1.29 & 0.18 & 0.35 & -0.69
& 0.21 & 0.15 & -0.44 & 0.48 & 0.03 & 2.18
& 0.17 & 0.88 \\ \hline
\end{tabular}
\caption{Parameter set that was calibrated in the eANN parameterization, which represents the last layer of the neural network (12 weights and 1 bias) and scaling coefficient in front of the parameterization ($\gamma$). Offline-trained values ("Offline") are obtained on the global ocean dataset CM2.6. The initial ensemble is drawn from the normal distribution, which has the offline-trained values as its mean and the standard deviation shown in the third row ("STD"). The weights share the same standard deviation at initialization, which is 25\% of the standard deviation over the parameter dimension. The standard deviation of the bias and the scaling coefficient is 25\% of their respective absolute values. Parameters of the calibrated parameterization in Double Gyre ("DG") and NeverWorld2 ("NW2") configurations are shown in the two bottom rows.}
\end{table}

\clearpage

\bibliography{agusample}